\documentclass[12pt,a4paper]{article}
\usepackage{times,epsfig,graphics}
\begin{document}
\title{Time-dependent embedding}
\author{J. E. Inglesfield,\\ School of Physics and Astronomy,
\\ Cardiff University, Cardiff, CF24 3YB, UK}
\date{}
\maketitle
\begin{abstract}
A method of solving the time-dependent Schr\"odinger equation is
presented, in which a finite region of space is treated explicitly,
with the boundary conditions for matching the wave-functions on to the
rest of space replaced by an embedding term added on to the
Hamiltonian. This time-dependent embedding term is derived from the
Fourier transform of the energy-dependent embedding potential, which
embeds the time-independent Schr\"odinger equation. Results are
presented for a one-dimensional model of an atom in a time-varying
electric field, the surface excitation of this model atom at a jellium
surface in an external electric field, and the surface excitation of a
bulk state.
\end{abstract}
\section{Introduction}
The availability of ultra-short laser pulses \cite{agost} opens up new
ways of studying time-dependent electronic processes in atoms
\cite{kien,qub}, molecules \cite{blanchet} and at solid surfaces
\cite{miaja}. In surface physics the time delay of a core electron 
emitted through the surface potential barrier can now be compared on
an attosecond timescale with valence electron photoemission in
time-resolved photoemission experiments \cite{caval}. This makes it
important to develop appropriate theoretical tools for studying such
processes, in particular for solving the time-dependent Schr\"odinger
equation accurately. A problem arises with the boundary in solving
this equation -- if an electron is ejected from an atom, for example,
how is its wave-function treated as it propagates outwards towards the
edge of the computational region? One approach is to apply absorbing
boundary conditions at the edge of the region \cite{kulander}, but
this is an approximation \cite{boucke}; complex coordinate methods can
also be used to remove the outgoing wave in solutions of the
time-dependent Schr\"odinger equation \cite{mccurdy}.  Other methods
are used in atomic physics to study photoionization in particular --
the wave-function can be expanded in complex basis functions
\cite{piraux}, and Floquet methods use a Floquet-Fourier series
expansion of the wave-function \cite{purvis}. 

In recent years there has been work on transparent boundary
conditions, so that the solution of the time-dependent Schr\"odinger
equation in some restricted region of interest, which we call
region I, propagates out through the boundary into the rest of the
system, region II, without reflection. It has been shown by Hellums
and Frensley \cite{hellums} using a matrix partitioning of the
Hamiltonian that these boundary conditions are equivalent to an extra
``memory'' term -- a time-dependent embedding or self-energy term --
added to the Hamiltonian. This spatial partitioning, and the
corresponding form of the embedding term, is appropriate for a
Hamiltonian constructed with localized basis functions or spatially
discretized wave-functions. 

An embedding method for solving the time-independent Schr\"odinger
equation in the region of interest was developed many years ago by the
author \cite{ingles1,ingles2}. The operator which embeds region I --
the embedding potential -- is given by the surface inverse of a Green
function for region II, evaluated over the boundary between the
regions. Any convenient basis set can be used to expand the
wave-functions (or Green function) in region I, and the method has
been widely used in accurate surface calculations \cite{ishida}, and
recently in photonics applications \cite{kemp}. The embedding
potential is a generalized logarithmic derivative, giving the normal
derivative of the wave-function over the boundary of region I in terms
of the amplitude -- in mathematical terms it is a Dirichlet-to-Neumann
map \cite{isakov}. It is this form of embedding potential, rather than
a tight-binding or discretized form, which we shall apply in this
paper to the time-dependent problem.

The time-dependent version of the Dirichlet-to-Neumann map has been
studied by Ehrhardt \cite{ehrhardt}, and applied to embedding the
time-dependent Schr\"odinger equation, discretized spatially as well
as temporally -- Ehrhardt paid particular attention to the stability
and accuracy of the time-evolution. Moyer \cite{moyer} has used this
to study a range of one-dimensional problems including the scattering
of electrons in model semiconductor structures. Recently Kurth
\emph{et al.} \cite{kurth} have developed a spatially discretized
method to study quantum transport through a structure between two
leads, replacing the leads by time-dependent embedding self-energies;
they also consider time-varying bias potentials in the leads. Boucke,
Schmitz and Kull \cite{boucke} have applied the time-dependent
relationship between normal derivative and amplitude to the
one-dimensional problem of an oscillating potential of the form
$V(z,t)=-1/\cosh^2[z+\xi_0\sin(\omega t)]$ acting on the wave-function
$1/\cosh(z)$. (This wave-function is the bound state of the static
potential; the oscillating potential is equivalent, by the
Kramers-Henneberger transformation \cite{hen}, to the static potential
in a spatially uniform time-varying electric field.) They solve this
problem on a finite spatial grid, with the boundary relationship, for
zero potential in the external region, applied to the end grid points.

Instead of using a finite-difference grid, Ermolaev \emph{et al.}
\cite{ermol} expand the wave-function in region I in a basis set. 
The matrix element of the kinetic energy operator evaluated over this
finite region of space gives a normal derivative term at the boundary
of region I, and this is evaluated using the Green function
relationship with the boundary amplitude. They also consider the
time-evolution of the bound state of the $-1/\cosh^2(z)$ potential,
but instead of transforming the spatially homogeneous, time-varying
electric field using the Kramers-Henneberger transformation, they
treat the problem directly, with a sinusoidally-varying field in
region II.  Like Ermolaev \emph{et al.} we will also use a basis set
for expanding the wave-function in region I, but we incorporate the
amplitude-derivative relationship in a way analogous to the
energy-dependent embedding method \cite{ingles1}.

We begin this paper with a derivation in section 2 of the
time-dependent embedding method, based on our original method for
embedding the stationary Schr\"odinger equation. In section 3 we
derive the time-dependent embedding potential analytically for a
constant external potential in region II, and demonstrate a numerical
technique for embedding on to a constant electric field. We move on to
model applications in sections 4 and 5. In section 4 we shall consider
the same problem as Boucke \emph{et al.} \cite{boucke} -- the
oscillating potential $-1/\cosh^2[z+\xi_0\sin(\omega t)]$ acting on
the $1/\cosh z$ bound state, using a very small region I enclosing
this potential and embedding on to zero potential on either side. We
shall see that the results are in excellent agreement with a
finite-difference calculation extending a large distance on either
side of the time-varying potential. In this section we shall next
consider the excitation of a localized wave-function at a the surface
of a free-electron metal by a time-dependent perturbation, with a
static electric field on the vacuum side of the surface. We could use
this in a model of photo-assisted field emission \cite{bagchi}, or
even pump-probe experiments; here the calculation provides a test of
the embedding potential for a uniform electric field.

The problems we solve in section 4 are restricted to wave-functions
confined initially to region I, though in the course of time-evolution
they leak outside. This is a severe restriction for surface physics or
other condensed matter applications, where the initial wave-function
is usually an extended state. In section 5 we shall show how the
time-evolution of such wave-functions can be calculated using
embedding, with region I containing the time-dependent perturbing
potential, but not necessarily the starting wave-function. The model
calculation we shall describe in section 5 corresponds to the surface
of a free-electron metal, with a sinusoidal time-dependent potential
applied in the surface region (region I). With a small basis set, the
time-evolution of the bulk wave-functions at the surface can be
calculated very accurately.

Atomic units are used throughout this paper, with $e=\hbar=m_e=1$. The
atomic unit of time $=2.418884\times 10^{-17}$ s, so that 1 fs 
$=41.34138$ a.u.

\section{Stationary and time-dependent embedding}
\begin{figure}[h]
\begin{center}
\epsfig{width=8cm,file=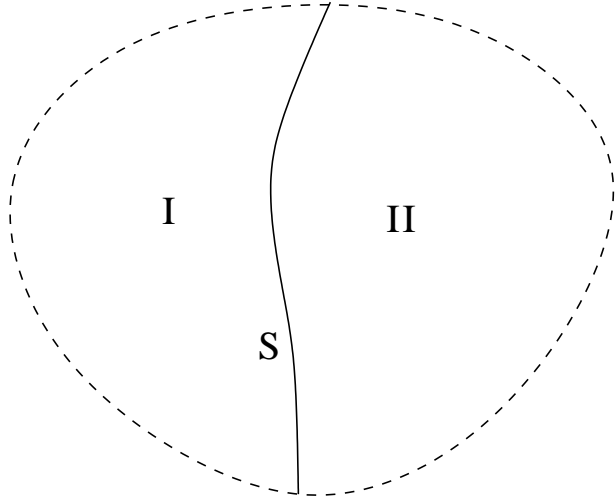}
\end{center}
\caption{Region I is embedded on to region II over surface S.}
\label{fig1}
\end{figure}

In this section we shall start from our previous results for embedding
the stationary Schr\"odinger equation \cite{ingles1}, and show how
these can be transformed into time-dependent embedding. The embedding
problem can be represented schematically by figure \ref{fig1}.
We wish to solve the Schr\"odinger equation -- either time-independent
or time-dependent -- in the whole system, regions I + II: region I is
treated explicitly, and region II is replaced by an ``embedding
potential'' at the interface S, added on to the Hamiltonian for region
I. In a typical application to surfaces, region I would be the surface
layer or two of atoms together with the potential barrier region, and
region II would be the vacuum on one side and the substrate crystal on
the other \cite{ishida}. The embedding potential ensures that the
wave-functions evaluated in region I match correctly in amplitude and
derivative on to the appropriate solutions of the Schr\"odinger
equation in region II.
\subsection{Stationary embedding}
The original embedding method is based on a variational principle for
the energy, and we shall here outline the derivation
\cite{ingles1,ingles2}. The one-electron Hamiltonian, of which we wish
to find the expectation value, is given by
\begin{equation}
H=-\frac{1}{2}\nabla^2+V(\mathbf{r}),
\label{emb0}
\end{equation}
where $V$ is the one-electron potential. In region I we have an
arbitrary trial function $\phi$; this is extended through region II
with the exact solution $\psi$ of the Schr\"odinger equation,
evaluated at some trial energy $\epsilon$, which matches in amplitude
on to $\phi$ over the boundary S joining the regions (figure
\ref{fig1}). It is assumed that at an external boundary (the dashed
line in figure \ref{fig1}), the wave-functions satisfy a homogeneous
boundary condition, typically going to zero. The expectation value of
$H$ is then given by
\begin{equation}
E=\frac{\int_I d\mathbf{r}\phi^*H\phi+\epsilon\int_{I\!I}d\mathbf{r}
\psi^*\psi+\frac{1}{2}\int_S d\mathbf{r}_S\phi^*\left(\frac{\partial
\phi}{\partial n_S}-\frac{\partial \psi}{\partial n_S}\right)}
{\int_I d\mathbf{r}\phi^*\phi+\int_{I\!I}d\mathbf{r}\psi^*\psi}.
\label{emb1}
\end{equation}
The first two terms in the numerator are the expectation values of the
Hamiltonian in regions I and II. The third term, an integral over the
boundary S, contains the difference in normal derivatives on either
side of S (measured outwards from I) and comes from the kinetic energy
operator acting on the kink in the trial function. 

We now use a result obtained by applying Green's theorem in region II,
\begin{equation}
\psi(\mathbf{r}_S)=-\frac{1}{2}\int_S d\mathbf{r}'_S G_0(\mathbf{r}_S,
\mathbf{r}'_S;\epsilon)\frac{\partial\psi(\mathbf{r}_S')}
{\partial n_S}
\label{emb2}
\end{equation}
-- $G_0$ is the Green function in II satisfying the zero normal
derivative boundary condition on S. Taking the inverse of (\ref{emb2})
gives the result which is central to the embedding method,
\begin{equation}
\frac{\partial\psi(\mathbf{r}_S)}{\partial n_S}=-2\int_S
d\mathbf{r}'_S G_0^{-1}(\mathbf{r}_S,\mathbf{r}'_S;\epsilon)
\psi(\mathbf{r}'_S),
\label{emb3}
\end{equation}
where $G_0^{-1}$ is the surface inverse over S.  $G_0^{-1}$ is the
embedding potential, and as we see from equation (\ref{emb3}) it is a
generalized logarithmic derivative. $G_0^{-1}$ is the same as the
mathematicians' Dirichlet-to-Neumann map
\cite{isakov,szymt}, mapping the Dirichlet boundary condition (the
amplitude specified over the boundary) on to the Neumann (the
derivative specified). Using this result, and the fact that
$\psi=\phi$ over the boundary, we obtain
\begin{equation}
\int_S d\mathbf{r}\phi^*\frac{\partial \psi}{\partial n_S}=
-2\int_S d\mathbf{r}_S\int_S d\mathbf{r}'_S \phi^*(\mathbf{r}_S)
G_0^{-1}(\mathbf{r}_S,\mathbf{r}_S';\epsilon)\phi(\mathbf{r}_S').
\label{emb4}
\end{equation}
To complete the simplification of equation (\ref{emb1}), we use a
second result involving the embedding potential \cite{ingles1},
\begin{equation}
\int_{I\!I} d\mathbf{r}\psi^*\psi=-\int_S d\mathbf{r}_S\int_S 
d\mathbf{r}'_S \phi^*(\mathbf{r}_S)
\frac{\partial G_0^{-1}(\mathbf{r}_S,\mathbf{r}_S';\epsilon)}
{\partial\epsilon}\phi(\mathbf{r}_S').
\label{emb5}
\end{equation}
Substituting (\ref{emb4}) and (\ref{emb5}) into (\ref{emb1}) gives us
the embedding variational principle,
\begin{equation}
E=\frac{\int_I d\mathbf{r}\phi^*H\phi
+\frac{1}{2}\int_S d\mathbf{r}_S\phi^*\frac{\partial
\phi}{\partial n_S}+\int_S d\mathbf{r}_S\int_S d\mathbf{r}'_S 
\phi^*\left(G_0^{-1}-\epsilon\frac{\partial G_0^{-1}}{\partial n_S}
\right)\phi}{\int_I d\mathbf{r}\phi^*\phi-\int_S d\mathbf{r}_S\int_S 
d\mathbf{r}'_S \phi^*(\mathbf{r}_S)\frac{\partial G_0^{-1}}
{\partial\epsilon}\phi}
\label{emb6}
\end{equation}
-- this gives $E$ in terms of the trial function $\phi$ defined in
region I and on its boundary S.

The wave-function $\phi$ which minimizes (\ref{emb6}) satisfies the
following equation in region I, including S,
\begin{eqnarray}
\lefteqn{\left(-\frac{1}{2}\nabla^2+V(\mathbf r)\right)\phi(\mathbf{r})
+}\nonumber\\
&&\delta(\mathbf{r}-\mathbf{r}_S)\left[\frac{1}{2}\right.
\frac{\partial\phi}{\partial n_S}+\left.\int_S d\mathbf{r}_S'
\left(G_0^{-1}(\epsilon)+
(E-\epsilon)\frac{\partial G_0^{-1}}{\partial \epsilon}\right)
\phi(\mathbf{r}'_S)\right]=E\phi(\mathbf{r}).\hspace{0.6cm}
\label{emb7}
\end{eqnarray}
The embedding potential evaluated at trial energy $\epsilon$ plus the
energy derivative term give $G_0^{-1}$ at the required energy $E$, to
first order in $(E-\epsilon)$. The surface terms in square brackets
vanish when $\phi$ has the correct normal derivative to match on to
the solution of the Schr\"odinger equation in region II, so this is
the correctly embedded solution of the Schr\"odinger equation.

To find the solutions of (\ref{emb6}) and (\ref{emb7}), we expand
$\phi$ in terms of basis functions $\chi_i$ (here we assume that they
are real and orthonormal),
\begin{equation}
\phi(\mathbf{r})=\sum_i a_i\chi_i(\mathbf{r}).
\label{emb8}
\end{equation} 
Substituting into (\ref{emb6}) and finding the stationary values of
$E$ gives the following embedded Schr\"odinger equation in matrix
form,
\begin{equation}
\sum_j(H_{ij}+\Sigma_{ij})a_j=Ea_i.
\label{emb9}
\end{equation}
The first term comes from the Hamiltonian and the surface derivative
term in the variational principle,
\begin{eqnarray}
H_{ij}&=&\int_I
d\mathbf{r}\chi_i(\mathbf{r})\left[-\frac{1}{2}\nabla^2+
V(\mathbf{r})\right]\chi_j(\mathbf{r})+\frac{1}{2}\int_S d\mathbf{r}_S
\chi_i(\mathbf{r}_S)\frac{\partial\chi_j(\mathbf{r}_S)}
{\partial n_S}\nonumber\\
&=&\frac{1}{2}\int_I d\mathbf{r}\nabla \chi_i(\mathbf{r})\cdot
\nabla \chi_j(\mathbf{r})+\int_I d\mathbf{r}\chi_i(\mathbf{r})
V(\mathbf{r})\chi_j(\mathbf{r}),
\label{emb10}
\end{eqnarray}
while the second term comes from the embedding potential,
\begin{equation}
\Sigma_{ij}=\int_S d\mathbf{r}_S\int_S d\mathbf{r}_S'\chi_i
(\mathbf{r}_S)\left[G_0^{-1}(\mathbf{r}_S,\mathbf{r}_S';\epsilon)+
(E-\epsilon)\frac{\partial G_0^{-1}(\mathbf{r}_S,\mathbf{r}_S';
\epsilon)}{\partial\epsilon}\right]\chi_j(\mathbf{r}_S').
\label{emb11}
\end{equation}
When $\epsilon$ lies in an energy continuum of region II, $G_0^{-1}$
is complex, so we usually find the Green function of $(H+\Sigma)$
rather than the eigenvectors.
\subsection{Time-dependent embedding}
We shall now build the time-dependent formalism on the results of the
last section. First we consider the the relationships between normal
derivative and amplitude as functions of time, corresponding to
(\ref{emb2}) and (\ref{emb3}) in the energy-dependent case. Taking the
Fourier transform of (\ref{emb2}) gives us
\begin{equation}
\tilde{\psi}(\mathbf{r}_S,t)=-\frac{1}{2}\int_S d\mathbf{r}_S
\int_{-\infty}^t dt'\tilde{G}_0(\mathbf{r}_S,
\mathbf{r}'_S;t-t')\frac{\partial\tilde{\psi}(\mathbf{r}_S',t')}
{\partial n_S}
\label{emb13}
\end{equation}
-- we use a tilde to indicate functions of time. $\tilde{G}_0$
satisfies the time-dependent Schr\"odinger equation in region II,
\begin{equation}
\left(-\frac{1}{2}\nabla^2+V(\mathbf{r})
-i\frac{\partial}{\partial t}\right)
\tilde{G}_0(\mathbf{r},\mathbf{r}';t-t')=\delta(\mathbf{r}-\mathbf{r}')
\delta(t-t'),
\label{emb14}
\end{equation}
with the zero-derivative boundary condition on S, and because we take
the retarded Green function we have,
\begin{equation}
\tilde{G}_0(\mathbf{r},\mathbf{r}';t-t')=0,\;\;t<t'
\label{emb15}
\end{equation} 
-- hence the upper limit of $t$ in the integral in (\ref{emb13}). This
can be derived directly by applying Green's theorem to the
inhomogeneous equation (\ref{emb14}), and the corresponding
homogeneous equation for $\tilde{\psi}(\mathbf{r},t)$ \cite{boucke}.

Turning to the inverse relation giving the derivative in terms of the
amplitude, we have to proceed differently, because the Fourier
transform of $G_0^{-1}$ does not converge -- it is an increasing
function of $\epsilon$. However, Ehrhardt \cite{ehrhardt} and Boucke
\emph{et al.} \cite{boucke} have shown that $\partial\tilde{\psi}(t)
/\partial n_S$ can be expressed in terms of the time-derivative of
$\tilde{\psi}(t)$ , and here we give a simple derivation. We re-write
the integrand in (\ref{emb3}) as
\begin{equation}
\int_S
d\mathbf{r}'_S\frac{G_0^{-1}(\mathbf{r}_S,\mathbf{r}'_S;\epsilon)}
{-i\epsilon}[-i\epsilon\psi(\mathbf{r}'_S,\epsilon)]
\label{emb16}
\end{equation}
-- the Fourier transform of $G_0^{-1}(\epsilon)/-i\epsilon$ converges,
and the transform of $-i\epsilon\psi(\epsilon)$ is
$\partial\tilde{\psi}(t)/\partial t$. Defining $\bar{G}_0^{-1}(t)$ as
the following Fourier transform,
\begin{equation}
\bar{G}_0^{-1}(\mathbf{r}_S,\mathbf{r}_S';t)=
\frac{1}{2\pi}\int_{-\infty}^{+\infty}d\epsilon\exp(-i\epsilon t)
\frac{G_0^{-1}(\mathbf{r}_S,\mathbf{r}'_S;\epsilon)}{-i\epsilon},
\label{emb17}
\end{equation}
the Dirichlet-to-Neumann equation in time becomes
\begin{equation}
\frac{\partial\tilde{\psi}(\mathbf{r}_S,t)}{\partial n_S}=-2
\int_S d\mathbf{r}_S\int_{-\infty}^t dt'\bar{G}_0^{-1}(\mathbf{r}_S,
\mathbf{r}'_S;t-t')\frac{\partial\tilde{\psi}(\mathbf{r}_S',t')}
{\partial t'}.
\label{emb18}
\end{equation}

We now turn to the embedding problem. The wave-function satisfying the
time-dependent equation, at this stage with a time-\emph{independent}
potential, can be written in region I in terms of solutions of the
embedded Schr\"odinger equation (\ref{emb7}),
\begin{equation}
\tilde{\phi}(\mathbf{r},t)=\sum_i a_i\phi_i(\mathbf{r})
e^{-i E_i t}.
\label{emb19}
\end{equation}
Equation (\ref{emb7}) simplifies if the
embedding potential is evaluated at the eigen-energy,
\begin{equation}
\left(-\frac{1}{2}\nabla^2+V(\mathbf{r})\right)\phi_i(\mathbf{r})+
\delta(\mathbf{r}-\mathbf{r}_S)\left[\frac{1}{2}
\frac{\partial\phi_i}{\partial n_S}+\int_S d\mathbf{r}_S'G_0^{-1}
(E_i)\phi_i(\mathbf{r}'_S)\right]=E_i\phi_i(\mathbf{r}),
\label{emb20}
\end{equation}
and multiplying this equation by the coefficients in (\ref{emb19}) and
summing over $i$ gives
\begin{eqnarray}
\lefteqn{\left(-\frac{1}{2}\nabla^2+V(\mathbf{r})\right)
\tilde{\phi}(\mathbf{r},t)+\delta(\mathbf{r}-\mathbf{r}_S)
\left[\frac{1}{2}\frac{\partial\tilde{\phi}}{\partial n_S}+\right.}
\nonumber\\&&\hspace{1cm}
\left.\int_S d\mathbf{r}_S'\sum_i a_ iG_0^{-1}(\mathbf{r}_S,
\mathbf{r}_S';E_i)\phi_i(\mathbf{r}_S')e^{-iE_it}\right]
=i\frac{\partial\tilde{\phi}}{\partial t}.
\label{emb21}
\end{eqnarray}
Using the same trick as in going from (\ref{emb16}) to (\ref{emb17}),
this becomes
\begin{eqnarray}
\lefteqn{\left(-\frac{1}{2}\nabla^2+V(\mathbf{r})\right)
\tilde{\phi}(\mathbf{r},t)+\delta(\mathbf{r}-\mathbf{r}_S)
\left[\frac{1}{2}\frac{\partial\tilde{\phi}}{\partial n_S}+\right.}
\nonumber\\&&\hspace{1cm}
\left.\int_S d\mathbf{r}_S'\int_{-\infty}^t dt'\bar{G}_0^{-1}
(\mathbf{r}_S,\mathbf{r}_S';t-t')\frac{\partial\tilde{\phi}
(\mathbf{r}_S',t')}{\partial t'}\right]
=i\frac{\partial\tilde{\phi}}{\partial t},
\label{emb22}
\end{eqnarray} 
an equation which holds for $\mathbf{r}$ inside region I and on the
boundary S.

To show how the embedding terms in (\ref{emb22}) work, we construct the
solution of the time-dependent equation in region II, $\tilde{\psi}$,
with the inhomogeneous boundary condition that it matches in amplitude
on to $\tilde{\phi}$ over S, at all times up to $t$,
\begin{equation}
\tilde{\psi}(\mathbf{r}_S,t')=\tilde{\phi}(\mathbf{r}_S,t'),\;t'\le t.
\label{emb23}
\end{equation}
From (\ref{emb18}), the normal derivative on S is given by
\begin{equation}
\frac{\partial\tilde{\psi}(\mathbf{r}_S,t)}{\partial n_S}=-2
\int_S d\mathbf{r}_S\int_{-\infty}^t dt'\bar{G}_0^{-1}(\mathbf{r}_S,
\mathbf{r}'_S;t-t')\frac{\partial\tilde{\phi}(\mathbf{r}_S',t')}
{\partial t'}.
\label{emb24}
\end{equation}
But the right-hand side is the second term in the square brackets in
(\ref{emb22}), with a factor of $-2$. For (\ref{emb22}) to be
satisfied on S as well as inside region I, the two terms in the square
brackets must cancel, forcing $\tilde{\phi}(\mathbf{r},t)$ to match in
normal derivative on to the solution in II; as they already match in
amplitude by construction, we have the correctly embedded solution of
the time-dependent equation. Moreover, the normal derivative term
combined with $-\frac{1}{2}\nabla^2$ give a Hermitian operator when
integrating over region I.

It is convenient to start off the time-evolution at $t=0$, assuming
that 
\begin{equation}
\tilde{\phi}(\mathbf{r}_S,t)=0,\;\;\;t<0,
\label{emb241}
\end{equation}
so we change the lower limit in the embedding term in (\ref{emb22})
from $-\infty$ to $t=0$. We also allow the potential in region I
to be a function of time -- we can presumably do this, as it does not
affect the surface integral, originating from region II. This gives us
the final form of the embedded time-dependent Schr\"odinger equation,
\begin{eqnarray}
\lefteqn{\left(-\frac{1}{2}\nabla^2+V(\mathbf{r},t)\right)
\tilde{\phi}(\mathbf{r},t)+\delta(\mathbf{r}-\mathbf{r}_S)
\left[\frac{1}{2}\frac{\partial\tilde{\phi}}{\partial n_S}+\right.}
\nonumber\\&&\hspace{1cm}
\left.\int_S d\mathbf{r}_S'\int_{0}^t dt'\bar{G}_0^{-1}
(\mathbf{r}_S,\mathbf{r}_S';t-t')\frac{\partial\tilde{\phi}
(\mathbf{r}_S',t')}{\partial t'}\right]
=i\frac{\partial\tilde{\phi}}{\partial t}.
\label{emb242}
\end{eqnarray}

As in the time-independent case, this equation can be solved using a
basis set expansion of $\tilde{\phi}(\mathbf{r},t)$,
\begin{equation}
\tilde{\phi}(\mathbf{r},t)=\sum_i a_i(t)\chi_i(\mathbf{r}).
\label{emb243}
\end{equation}
Substituting into (\ref{emb242}) we obtain the matrix form of the
embedded Schr\"odinger equation,
\begin{equation}
\sum_j\left(H_{ij}(t)a_j(t)+\int_0^tdt'\bar{\Sigma}_{ij}(t-t')
\frac{da_j}{dt'}\right)=i\frac{da_i}{dt},
\label{emb25}
\end{equation}
with the Hamiltonian matrix given by
\begin{equation}
H_{ij}(t)=\frac{1}{2}\int_I d\mathbf{r}\nabla \chi_i(\mathbf{r})\cdot
\nabla \chi_j(\mathbf{r})+\int_I d\mathbf{r}\chi_i(\mathbf{r})
V(\mathbf{r},t)\chi_j(\mathbf{r}),
\label{emb26}
\end{equation}
and the embedding matrix by
\begin{equation}
\bar{\Sigma}_{ij}(t)=\int_S d\mathbf{r}_S\int_S d\mathbf{r}_S'
\chi_i(\mathbf{r}_S)\bar{G}_0^{-1}(\mathbf{r}_S,\mathbf{r}'_S;t)
\chi_j(\mathbf{r}_S').
\label{emb27}
\end{equation}
The structure of (\ref{emb242}) and (\ref{emb25}) is the same as in
the spatially discretized approach to the problem
\cite{hellums,kurth}, with an embedding operator added on to the
time-dependent Schr\"odinger equation.

\section{The time-dependent embedding kernel}
In this section we shall evaluate the embedding kernel
$\bar{G}_0^{-1}(t)$ for constant and linear potentials in region
II. We work in one-dimension, but we shall see in section 3.3 how the
results can be applied to three-dimensional problems.
\subsection{Embedding on to zero potential}
In the one-dimensional case we use (\ref{emb3}) to determine
$G_0^{-1}(\epsilon)$, and then substitute into the Fourier transform
(\ref{emb17}) to find $\bar{G}_0^{-1}(t)$. With zero potential the
wave-functions in region II satisfying outgoing boundary conditions
for $z\ge0$ are
\begin{equation}
\psi(z,\epsilon)=\left\{\begin{array}{l}
\exp(-\gamma z),\;\;\gamma=\sqrt{-2\epsilon},\;\;
\epsilon<0\\
\exp(i\kappa z),\;\;\kappa=\sqrt{2\epsilon},\;\;\epsilon>0.
\end{array}\right.
\label{td1}
\end{equation}
Hence the embedding potential is 
\begin{equation}
G_0^{-1}(\epsilon)=\sqrt{-\epsilon/2}\;\;\mbox{or}\;\; 
-i\sqrt{\epsilon/2},
\label{td2}
\end{equation}
and substituting into (\ref{emb17}) the time-dependent embedding
kernel is given by 
\begin{equation}
\bar{G}_0^{-1}(t)=\frac{1}{2\pi}\int_{-\infty}^{+\infty}d\epsilon
\exp(-i\epsilon t)\left\{\begin{array}{c}\frac{-i}{\sqrt{-2\epsilon}},
\;\epsilon<0\\ \frac{1}{\sqrt{2\epsilon}},\;\epsilon>0.\end{array}
\right.
\label{td3}
\end{equation}
This integral can be done analytically \cite{boucke}, and the result
is
\begin{equation}
\bar{G}_0^{-1}(t)=\left\{\begin{array}{l}
0,\;\;t<0\\\frac{1-i}{2\sqrt\pi}\cdot\frac{1}{\sqrt{t}},\;\;t>0.
\end{array}\right.
\label{td4}
\end{equation}
We shall apply this embedding kernel in applications in sections 4 and
5.
\subsection{Embedding on to an electric field}
With other one-dimensional potentials it is necessary to carry out the
Fourier transform in (\ref{emb17}) numerically, and we consider the
linear potential corresponding to a constant electric field
$\mathcal{E}$. The wave-functions in region II satisfy the
Schr\"odinger equation,
\begin{equation}
-\frac{1}{2}\frac{d^2\psi}{dz^2}-\mathcal{E}z\psi=\epsilon\psi,
\label{td5}
\end{equation} 
which has Airy function solutions \cite{abram},
\begin{eqnarray}
\psi_1(z)&=&\mbox{Ai}\left(-(2\mathcal{E})^{\frac{1}{3}}(z+\epsilon/
\mathcal{E})\right),\nonumber\\
\mbox{and}\;\; \psi_2(z)&=&\mbox{Bi}\left(-(2\mathcal{E})
^{\frac{1}{3}}(z+\epsilon/\mathcal{E})\right).
\label{td6}
\end{eqnarray}
Now we need the combination of $\psi_1$ and $\psi_2$ corresponding to
outgoing waves, and from the asymptotic behaviour of the Airy
functions \cite{abram}, this is given by
\begin{equation}
\psi_+(z)=\mbox{Bi}\left(-(2\mathcal{E})^{\frac{1}{3}}(z+\epsilon/
\mathcal{E})\right)+i\mbox{Ai}\left(-(2\mathcal{E})^{\frac{1}{3}}
(z+\epsilon/\mathcal{E})\right).
\label{td7}
\end{equation}
So in this case the energy-dependent embedding potential at $z=0$ is
\begin{equation}
G_0^{-1}(\epsilon)=\frac{(2\mathcal{E})^{\frac{1}{3}}\left[
\mbox{Bi}'\left(-(2\mathcal{E})^{\frac{1}{3}}\epsilon/\mathcal{E}\right)
+i\mbox{Ai}'\left(-(2\mathcal{E})^{\frac{1}{3}}\epsilon/
\mathcal{E}\right)\right]}{2\left[\mbox{Bi}\left(-(2\mathcal{E})
^{\frac{1}{3}}\epsilon/\mathcal{E}\right)+i\mbox{Ai}
\left(-(2\mathcal{E})^{\frac{1}{3}}\epsilon/\mathcal{E}\right)\right]},
\label{td8}
\end{equation}
which we evaluate using the Airy function programs due to Gil \emph{et
al.} \cite{gil}.  For large positive or negative $\epsilon$,
$G_0^{-1}(\epsilon)$ has the $\sqrt{\epsilon}$ free-electron behaviour
given by equation (\ref{td2}), so once again we divide by $\epsilon$
when finding the Fourier transform to obtain the time-dependent
kernel. The free-electron behaviour at large $\epsilon$ is what we
would expect -- the potential becomes irrelevant in this limit.

In the numerical Fourier transform of $G_0^{-1}(\epsilon)/\epsilon$ we
must be careful about the singularity at $\epsilon=0$, and we rewrite
the integral as
\begin{eqnarray}
\lefteqn{\bar{G}_0^{-1}(t)=\frac{i}{2\pi}\int_{-\infty}^{+\infty}
d\epsilon
\exp(-i\epsilon t)\frac{G_0^{-1}(\epsilon)}{\epsilon}}\nonumber\\ 
&=&\frac{i}{2\pi}\int_{-\infty}^{+\infty}d\epsilon
\exp(-i\epsilon t)\frac{G_0^{-1}(\epsilon)-G_0^{-1}(0)}{\epsilon}
+\frac{iG_0^{-1}(0)}{2\pi}\int_{-\infty}^{+\infty}d\epsilon
\frac{\exp(-i\epsilon t)}{\epsilon}.\hspace{1cm}
\label{td9}
\end{eqnarray}
The first integral is well-behaved at $\epsilon=0$, but in evaluating
the second integral we have to be careful about the contour of
integration around this point. As $\bar{G}_0^{-1}(t)$ is zero for
$t<0$, the singularities of the integrand must lie in the lower
half-plane, so we replace $\frac{1}{\epsilon}$ by
$\frac{1}{\epsilon+i\eta}$, where $\eta$ is infinitesimal. Our Fourier
transform then becomes
\begin{equation}
\bar{G}_0^{-1}(t)=\frac{i}{2\pi}\int_{-\infty}^{+\infty}d\epsilon
\exp(-i\epsilon t)\frac{G_0^{-1}(\epsilon)-G_0^{-1}(0)}{\epsilon}
+\left\{\begin{array}{l}0,\;\;t<0\\G_0^{-1}(0),\;\;t>0.
\end{array}\right.
\label{td11}
\end{equation}
To evaluate the integral in (\ref{td11}), our procedure is to include
an exponential damping term $\exp(-|\epsilon|/\Gamma)$, and discretize
the integral with finite limits. This simple method of evaluating the
Fourier transform works well.

\begin{figure}[h]
\begin{center}
\epsfig{width=8cm,angle=-90,file=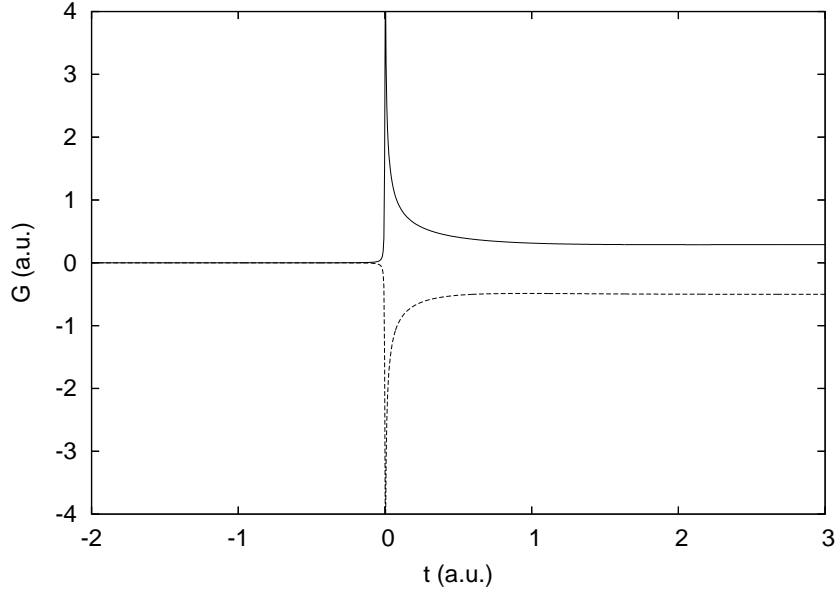}
\end{center}
\caption{$\bar{G}_0^{-1}(t)$ for the Schr\"odinger equation with 
electric field $\mathcal{E}=2$ a.u. Solid line,
$\mathcal{R}\mbox{e}\bar{G}_0^{-1}$; dashed line,
$\mathcal{I}\mbox{m}\bar{G}_0^{-1}$.}
\label{fig2}
\end{figure}

Results for the time-dependent embedding kernel with an electric field
$\mathcal{E}=2$ a.u. are shown in figure \ref{fig2}, using a
coefficient $\Gamma=500$ a.u. in the damping term. We see that
$\bar{G}_0^{-1}(t)$ is accurately zero for negative $t$. For
$t\rightarrow 0$ from above, the kernel tends to the free-electron
value (\ref{td4}) in accordance with our intuition: at short times (or
high energies) an electron cannot tell whether it is in an electric
field or not. For larger $t$, $\bar{G}_0^{-1}(t)$ rapidly approaches
the zero-frequency embedding potential, $G_0^{-1}(0)$, which is the
contribution from the pole in the frequency integral (\ref{td9}); the
larger the field, the more rapidly it approaches a constant
value. This form of $\bar{G}_0^{-1}(t)$ makes it very easy to evaluate
the long-time contribution to the embedded Schr\"odinger equation.

We test the accuracy of the embedding kernel by calculating the
time-evolution of a wave-function $\tilde{\psi}(z,t)$ over the
extended system of regions I and II, and then use (\ref{emb18}) to
compare the Dirichlet-to-Neumann result with the directly-calculated
derivative at the boundary of region II. In this test the potential
itself is time-independent, with an infinite barrier at $z=0$, zero
potential between $z=0$ and $z'$, and in region II beyond $z'$ an
electric field corresponding to the potential
$-\mathcal{E}(z-z')$. The wave-function at $t=0$ is taken to be a sum
of Gaussians vanishing at $z=0$,
\begin{equation}
\tilde{\psi}(z,0)=\exp\left[-\left(\frac{z-z_0}{w}\right)^2\right]
-\exp\left[-\left(\frac{z+z_0}{w}\right)^2\right],\;\;z>0,
\label{td12}
\end{equation}
and the value of $z_0$ and the width $w$ are chosen so that the
initial value of the wave-function at $z'$, where the electric field
starts, is negligible. To solve the time-dependent Schr\"odinger
equation for $\tilde{\psi}(z,t)$ we discretize in both space and time,
evolving forward in time steps $\delta t$ using the norm-conserving
Crank-Nicolson method \cite{press},
\begin{equation}
\tilde{\psi}(z,t+\delta t)=\left(1+\frac{i\delta t H}{2}\right)^{-1}
\left(1-\frac{i\delta t H}{2}\right)\tilde{\psi}(z,t),
\label{td13}
\end{equation}
where $H$ is the finite-difference Hamiltonian matrix \cite{jos}.

\begin{figure}[h]
\begin{center}
\epsfig{width=14cm,file=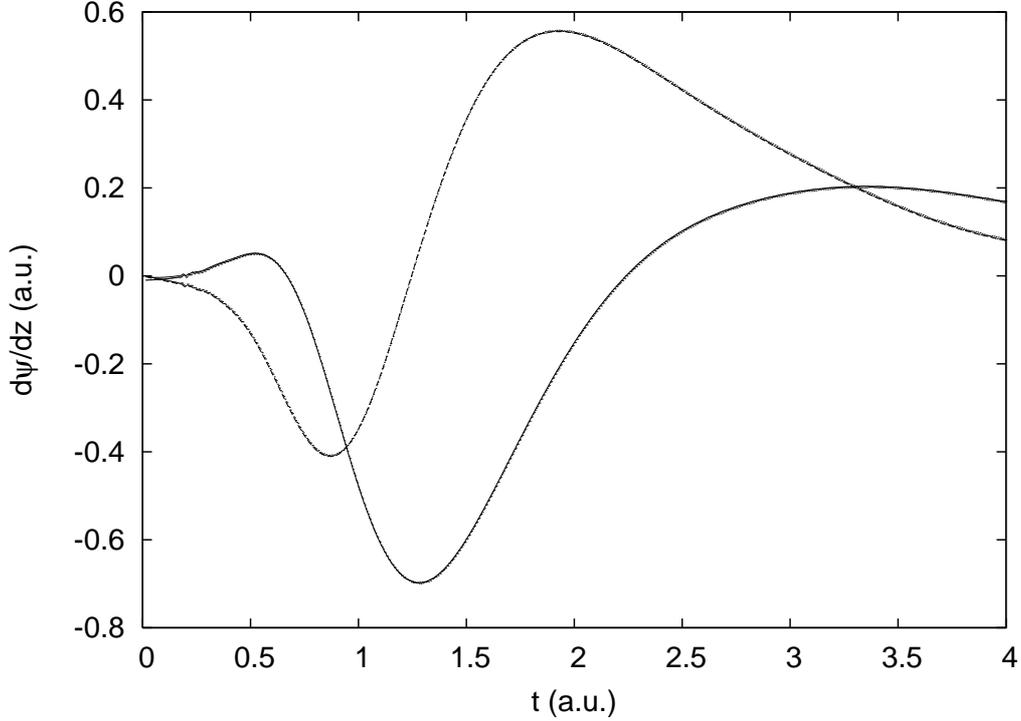}
\end{center}
\caption{Electric field test: $\partial\tilde{\psi}/\partial z$ 
at $z'$ evaluated from the Dirichlet-to-Neumann relation compared with
direct calculation from the numerical wave-function. D-to-N results
are the solid line (real part) and the dashed line (imaginary part);
direct results are the dots, hardly distinguishable except close to
$t=0$.}
\label{fig3}
\end{figure}

For this test we take an electric field $\mathcal{E}=2$ a.u. beyond
$z'=3$ a.u., with the constants in the Gaussian wave-function
$z_0=0.5$ a.u., $w=1$ a.u. The intervals in the spatial and time
discretization are $\delta z=0.0005$ a.u.  and $\delta t=0.005$ a.u.,
and the spatial range extends to $z=40$ a.u. -- enough to eliminate
edge effects over the time-range we consider. We then evaluate
$\partial\tilde{\psi}/\partial z$ at $z'$ directly from the finite
difference results, using first order differences, and compare this
with the derivative evaluated using (\ref{emb18}) with the embedding
kernel shown in figure \ref{fig2}, and first-order differences for
$\partial\tilde{\psi}/\partial t$. The comparison is shown in figure
\ref{fig3}: the values of $\partial\tilde{\psi}/\partial z$ from the
Dirichlet-to-Neumann relationship (solid and dashed lines) can hardly
be distinguished from the results for the derivative calculated
directly (dots). Only at very short times, and because of some noise
in the direct results, can they be distinguished. This shows that our
numerical evaluation of the embedding kernel for the electric field is
accurate -- the method can presumably be extended to other potentials.
\subsection{Shifting the potential}
If we know the embedding kernel for a one-dimensional potential
$V(\mathbf{r})$ in region II, we can immediately find an expression
analogous to the Dirichlet-to-Neumann relationship (\ref{emb18}) for
the potential shifted by a constant $V_0$. The time-dependent
Schr\"odinger equation satisfied by $\tilde{\psi}(\mathbf{r},t)$ in
region II is given by
\begin{equation}
\left(-\frac{1}{2}\nabla^2+V(\mathbf{r})+V_0\right)
\tilde{\psi}(\mathbf{r},t)=i\frac{\partial\tilde{\psi}}{\partial t},
\label{td15}
\end{equation}
and making the substitution
\begin{equation}
\tilde{\psi}(\mathbf{r},t)=\exp(-iV_0 t)\tilde{\Psi}(\mathbf{r},t),
\label{td16}
\end{equation}
$\tilde{\Psi}$ satisfies the unshifted Schr\"odinger equation
\begin{equation}
\left(-\frac{1}{2}\nabla^2+V(\mathbf{r})\right)
\tilde{\Psi}(\mathbf{r},t)=i\frac{\partial\tilde{\Psi}}{\partial t}.
\label{td17}
\end{equation}
As $\partial\tilde{\Psi}/\partial n_S$ is related to
$\partial\tilde{\Psi}/\partial t$ by the original embedding kernel
for $V(\mathbf{r})$, the corresponding relationship for
$\tilde{\psi}$ with the shifted potential is given by
\begin{eqnarray}
\lefteqn{\frac{\partial\tilde{\psi}(\mathbf{r}_S,t)}{\partial n_S}=
-2\exp(-iV_0t)}\hspace{12cm}\nonumber \\
\times\int_S d\mathbf{r}_S\int_{0}^t dt'
\bar{G}_0^{-1}(\mathbf{r}_S,\mathbf{r}'_S;t-t')\frac{\partial}
{\partial t'}[\exp(iV_0t')\tilde{\psi}(\mathbf{r}_S',t')].
\label{td18}
\end{eqnarray}

This result will be particularly useful when we come to deal with the
potential step at a surface. Moreover, a one-dimensional potential has
often been used in surface calculations \cite{chulkov}, and then the
solutions of the Schr\"odinger equation have the form
\begin{equation}
\tilde{\psi}(\mathbf{r},t)=\exp(i\mathbf{K}\cdot\mathbf{R})
\hat{\psi}_K(z,t),
\label{td19}
\end{equation}
where $\mathbf{K}$ is the Bloch wave-vector parallel to the surface,
and $\mathbf{R}$ is the surface-parallel component of
$\mathbf{r}$. Then $\hat{\psi}_K$ satisfies the one-dimensional
Schr\"odinger equation shifted by $K^2/2$, and the
Dirichlet-to-Neumann relationship becomes
\begin{equation} \frac{\partial\hat{\psi}_K(z,t)}{\partial z}=
-2\exp\left(-\frac{i K^2 t}{2}\right)\int_{0}^t dt'
\bar{G}_0^{-1}(t-t')\frac{\partial}{\partial t'}\left[\exp\left(
\frac{i K^2 t'}{2}\right)\hat{\psi}_K(z,t')\right],
\label{td20}
\end{equation} 
where $\bar{G}_0^{-1}$ is the embedding kernel for the one-dimensional
potential. The right-hand side of (\ref{td20}) provides the embedding
potential for states with Bloch wavevector $\mathbf{K}$ in this form
of potential. 

Another approximation for surfaces is to assume the full
three-dimensional potential for the surface region itself (region I),
and a one-dimensional potential for the bulk substrate and the vacuum
(which together constitute region II) \cite{ishida_liebsch}. Again
(\ref{td20}) can be used to construct the embedding potential for each
Fourier component of the wave-function in region I. We shall explore
this in a later paper.

\section{Embedding}
\subsection{Model atomic problem}
\begin{figure}[h]
\begin{center}
\epsfig{width=12cm,file=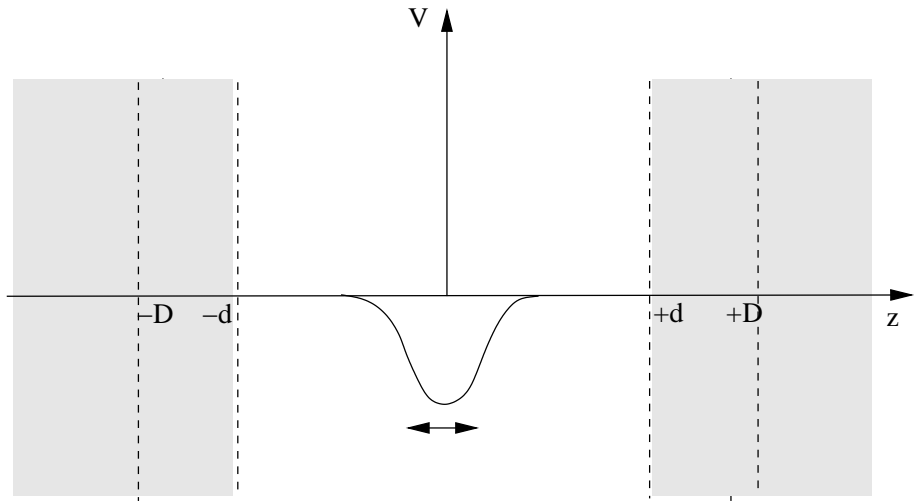}
\end{center}
\caption{Oscillating model atomic potential. Region I, treated
explicitly, lies within $z=\pm d$. The basis functions for expanding
the wave-function in region I, are defined in terms of $z=\pm D$.}
\label{fig4}
\end{figure}
To test our time-dependent embedding method with the kernels derived
in section 3, in this section we calculate the time-evolution of
states which are initially localized in region I. We start off with
the time-evolution of the normalized bound state wave-function
$\psi(z)=1/\sqrt 2\cosh(z)$ of the one-dimensional potential
$V(z)=-1/\cosh^2(z)$ in a time-varying electric field
$\mathcal{E}=\mathcal{E}_0\sin\omega t$. Ermolaev \emph{et al.}
\cite{ermol} solve this problem directly, but we
follow Boucke \emph{et al.} \cite{boucke}, who use the
Kramers-Henneberger transformation \cite{hen} to convert this into the
problem of the wave-function evolving in the oscillating potential,
\begin{equation}
V(z,t)=-1/\cosh^2[z+\xi_0\sin(\omega t)]
\label{try1}
\end{equation}
with zero potential beyond (figure \ref{fig4}). Here $\xi_0$ is the
classical amplitude of oscillation in the electric field,
$\xi_0=\mathcal{E}_0/\omega^2$.  We solve the embedded Schr\"odinger
equation (\ref{emb242}, \ref{emb25}) in region I, defined as $|z|<d$,
replacing the regions with $|z|>d$ by the zero-potential embedding
kernel. The time-dependent wave-function in region I,
$\tilde{\phi}(z,t)$, is expanded in a basis set given by
\begin{equation}
\Xi_m=\left\{\begin{array}{l}\cos\frac{m\pi z}{2D},\,
\;m\mbox{ even}\\\sin\frac{m\pi z}{2D},\;\;m\mbox{ odd}\end{array}
\right.,
\label{try2}
\end{equation}
where $D$ lies beyond $d$ (figure \ref{fig4}) to give flexibility in
amplitude and derivative at the boundary of region I. We orthogonalize
and normalize the basis functions within region I by diagonalizing the
overlap matrix $S_{mn}$,
\begin{equation}
S_{mn}=\int_{-d}^{+d}dz\;\Xi_m(z)\Xi_n(z).
\label{try3}
\end{equation}
Defining $\alpha^j$ as the $j$'th eigenvector of $S$, with eigenvalue
$s_j$, the $j$'th orthonormalized basis function is given by
\begin{equation}
\chi_j(z)=\frac{1}{\sqrt s_j}\sum_m\alpha_m^j\,\Xi_m(z).
\label{try4}
\end{equation}
If overcompleteness is a problem, this will show itself as a very
small value of $s_j$, and the corresponding basis function can be
dropped. 

The time-dependent matrix equation (\ref{emb25}) becomes, in this
one-dimensional case with embedding at both ends of the range,
\begin{eqnarray}
\lefteqn{\sum_j H_{ij}(t)a_j(t)+\chi_i(-d)\int_0^t dt'
\bar{G}_{0}^{-1}(t-t')\frac{\partial\tilde{\phi}(-d,t')}{\partial t'}}
\hspace{3cm}\nonumber\\
&&+\chi_i(d)\int_0^t dt'\bar{G}_{0}^{-1}(t-t')\frac{\partial\tilde{\phi}
(d,t')}{\partial t'}=i\frac{d a_i}{dt},
\label{try5}
\end{eqnarray}
where the Hamiltonian matrix element is given by
\begin{equation}
H_{ij}(t)=\int_{-d}^{+d}dz \left(\frac{1}{2}\frac{d\chi_i}{dz}
\frac{d\chi_j}{dz}+\chi_i(z)V(z,t)\chi_j(z)\right),
\label{try6}
\end{equation}
and the embedding kernel $\bar{G}_{0}^{-1}$ by (\ref{td4}).

We turn to the numerical time-integration of (\ref{try5}), which we
write as 
\begin{equation}
\frac{da}{dt}+iHa=-i\Gamma,
\label{try7}
\end{equation}
where $\Gamma$ is the vector representing the embedding terms in
(\ref{try5}),
\begin{equation}
\Gamma_i=\chi_i(-d)\int_0^t dt'\bar{G}_{l}^{-1}(t-t')
\frac{\partial\tilde{\phi}(-d,t')}{\partial t'}
+\chi_i(d)\int_0^t dt'\bar{G}_{r}^{-1}(t-t')\frac{\partial\tilde{\phi}
(d,t')}{\partial t'}.
\label{try8}
\end{equation}
We can improve on the first-order integration scheme,
\begin{equation}
a(t+\delta t)=[1+i\delta t H(t)]^{-1}[a(t)-i\delta t\Gamma(t)],
\label{try10}
\end{equation}
by expanding the time-evolution operator to second order in $\delta
t$, giving
\begin{equation}
a(t+\delta t)=\left(1+i\delta t H(t)-\frac{\delta t^2 H(t)^2}{2}
\right)^{-1}[a(t)-i\delta t\Gamma(t)].
\label{try12}
\end{equation}
Although it is not consistently second order in $\delta t$, this
stable scheme proves more accurate than (\ref{try10}) in our tests.
In evaluating the integrals in (\ref{try8}) we use
\begin{equation}
\frac{\partial\tilde{\phi}(t)}{\partial t}\approx
\frac{\tilde{\phi}(t)-\tilde{\phi}(t-\delta t)}{\delta t},
\label{try13}
\end{equation}
rather than the more accurate formula 
\begin{equation}
\frac{\partial\tilde{\phi}(t)}{\partial t}\approx 
\frac{\tilde{\phi}(t+\delta t)-\tilde{\phi}(t-\delta t)}{2\delta t},
\label{try14} 
\end{equation}
because (\ref{try14}) cannot be applied at the upper limit of the
integral -- we do not yet know $\tilde{\phi}(t+\delta t)$. To do the
integration itself we subtract off the $(t-t')^{-1/2}$ singularity at
$t'=t$, and then use the trapezium rule for the remaining well-behaved
integral.

We calculate $\tilde{\phi}(z,t)$ in the oscillating potential
(\ref{try1}), with an amplitude of oscillation $\xi_0=2.5$ a.u. and
frequency $\omega=0.2$ a.u., corresponding to an electric field
$\mathcal{E}_0=0.1$ a.u. (These are the values used by Boucke \emph{et
al.} \cite{boucke} -- as the bound state energy of the starting
wave-function is $-0.5$ a.u., ionization is due to multiphoton
processes.) Region I is taken with $d=10$ a.u., and the basis
functions are defined with $D=13$ a.u.; the results presented below
are for 25 and 40 basis functions. An interval $\delta t =0.01$
a.u. is used in the time integration (\ref{try12}). As a benchmark we
compare the embedding results with the wave-function
$\tilde{\psi}(z,t)$ calculated over an extended range, using finite
differences and Crank-Nicolson time-integration (\ref{td13})
(throughout this section we use $\tilde{\psi}(z,t)$ to indicate the
wave-function through space extended beyond region I). The extended
wave-function is calculated over the range $-400<z<400$ a.u., with a
spatial interval $\delta z=0.004$ a.u., and a time-interval $\delta
t=0.02$ a.u.

\begin{figure}
\centering
\includegraphics*[width=12cm]{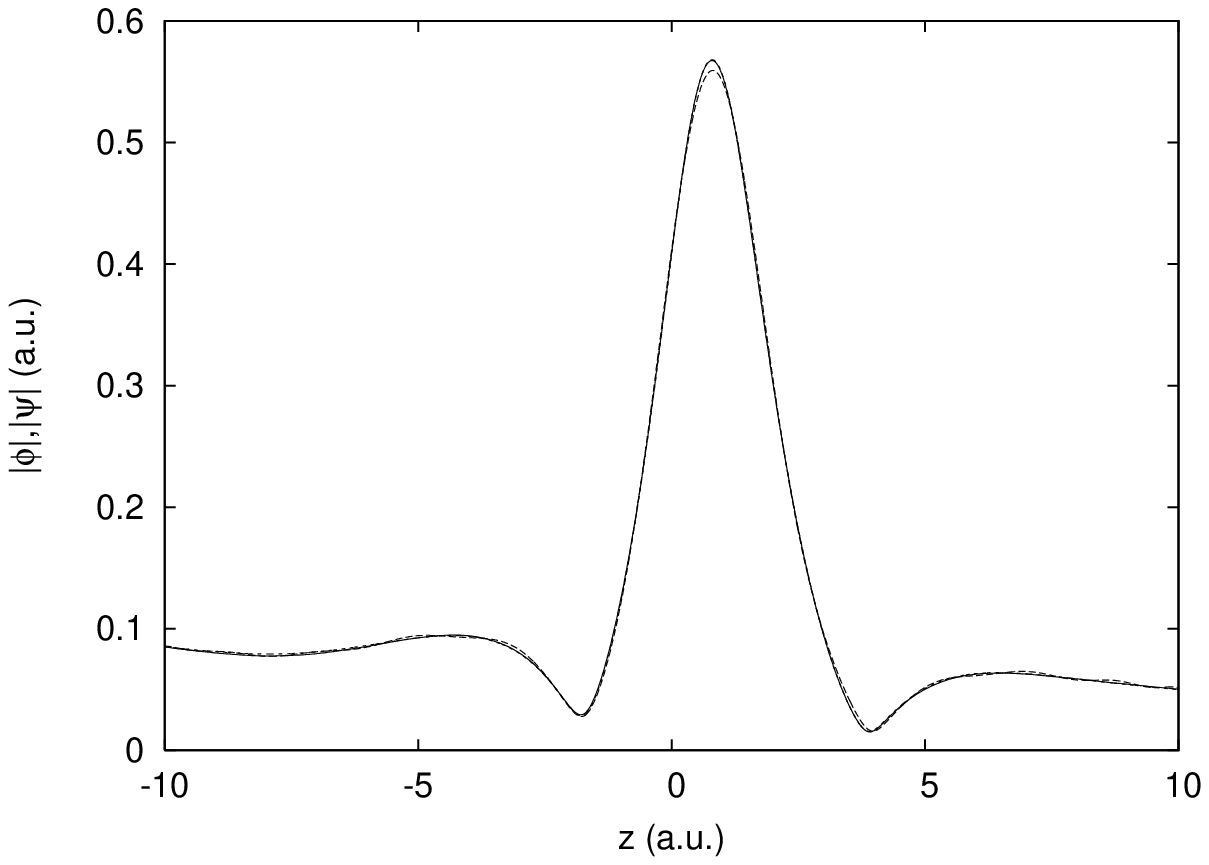}(a)
\includegraphics*[width=12cm]{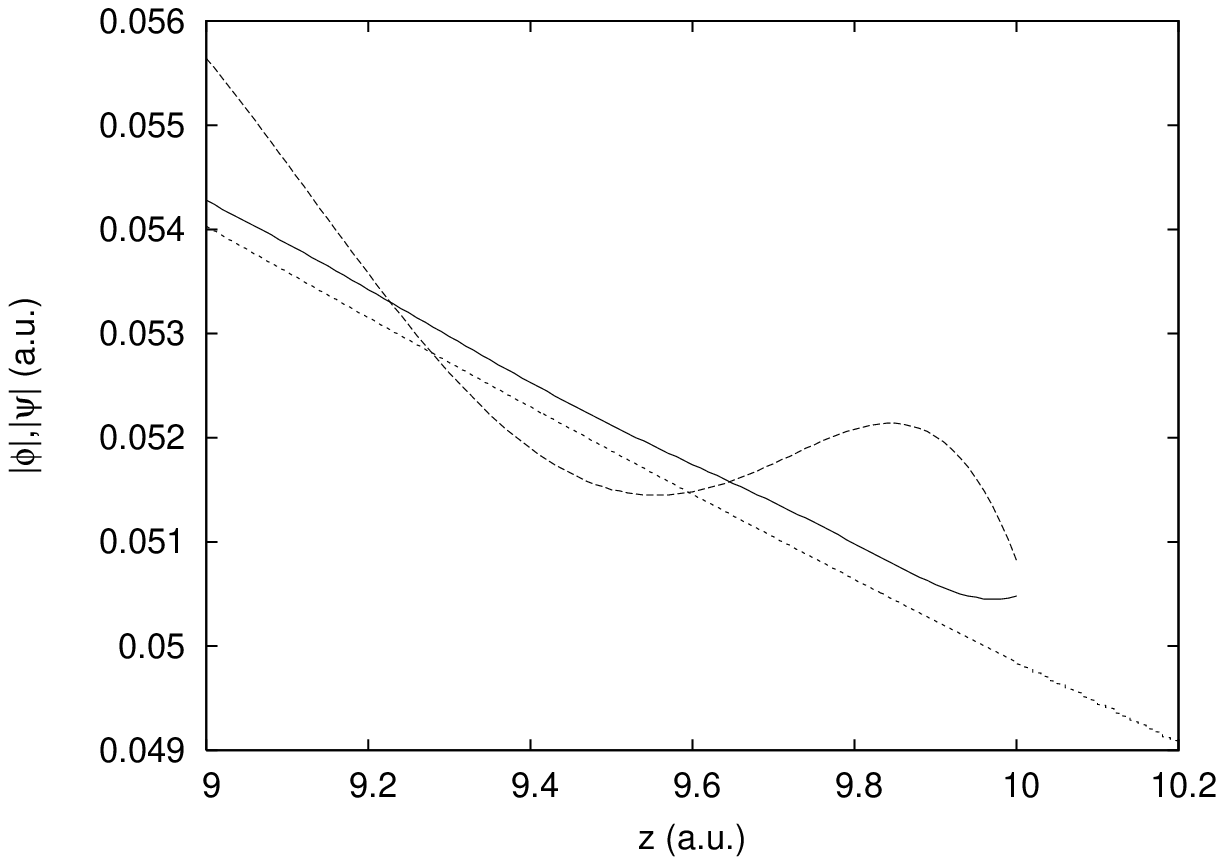}(b)
\caption{Oscillating potential: magnitude of wave-function at $t=80$ 
a.u., 2.55 periods.  $|\tilde{\phi}|$ calculated using embedding:
solid line, 40 basis functions; dashed line, 25 basis
functions. $|\tilde{\psi}|$ calculated over extended space using
finite differences: short-dashed line.\newline(a) Plotted over
embedding region between $z=\pm 10$ a.u. $|\tilde{\phi}|$ with 40
basis functions and $|\tilde{\psi}|$ are indistinguishable on this
scale.\newline(b) Plotted around the embedding point at $z=10$ a.u.}
\label{fig5}
\end{figure}
\begin{figure}
\centering
\includegraphics*[width=12cm]{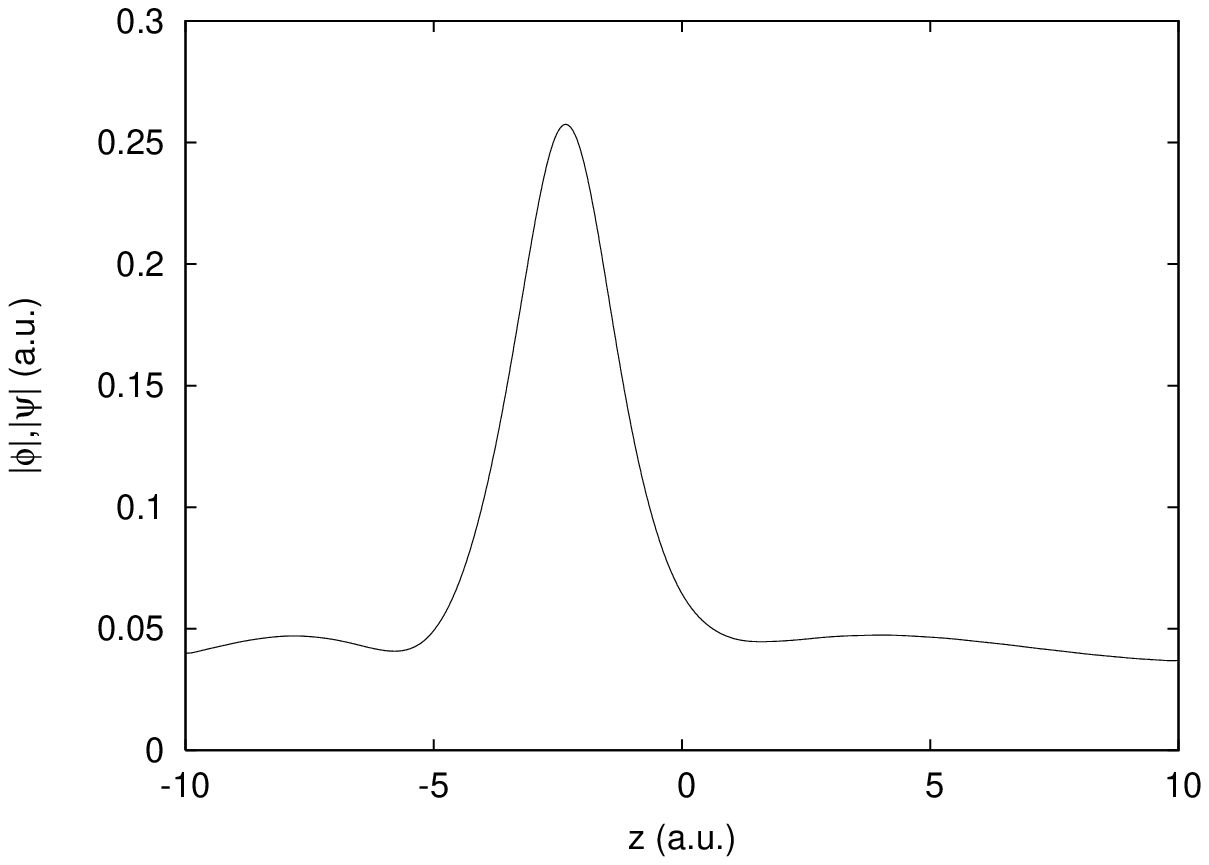}(a)
\includegraphics*[width=12cm]{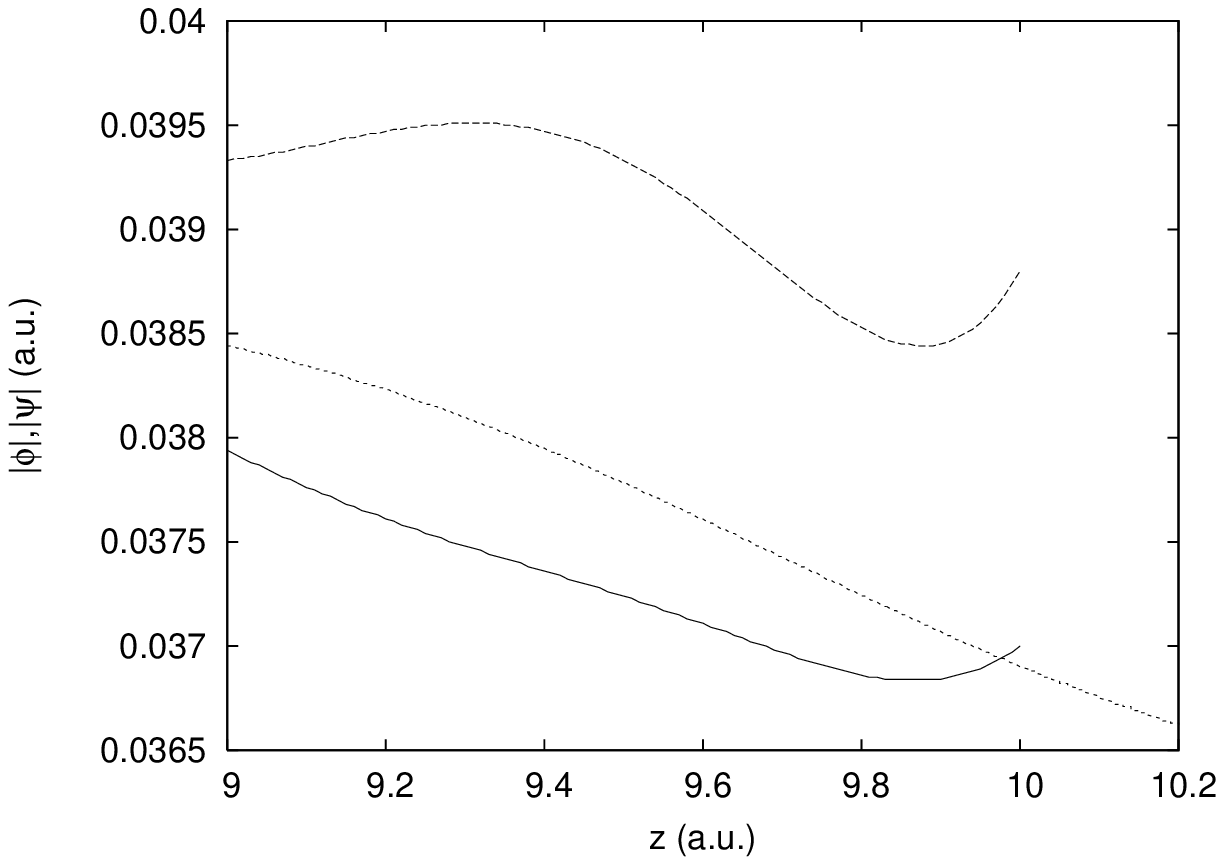}(b)
\caption{Oscillating potential: magnitude of wave-function at $t=320$ 
a.u., 10.19 periods.  $|\tilde{\phi}|$ calculated using embedding:
solid line, 40 basis functions; dashed line, 25 basis
functions. $|\tilde{\psi}|$ calculated over extended space using
finite differences: short-dashed line.\newline(a) Plotted over
embedding region between $z=\pm 10$ a.u. $|\tilde{\phi}|$ with 40
basis functions and $|\tilde{\psi}|$ are indistinguishable on this
scale.\newline(b) Plotted around the embedding point at $z=10$ a.u.}
\label{fig6}
\end{figure}
\begin{figure}
\centering
\includegraphics*[width=12cm]{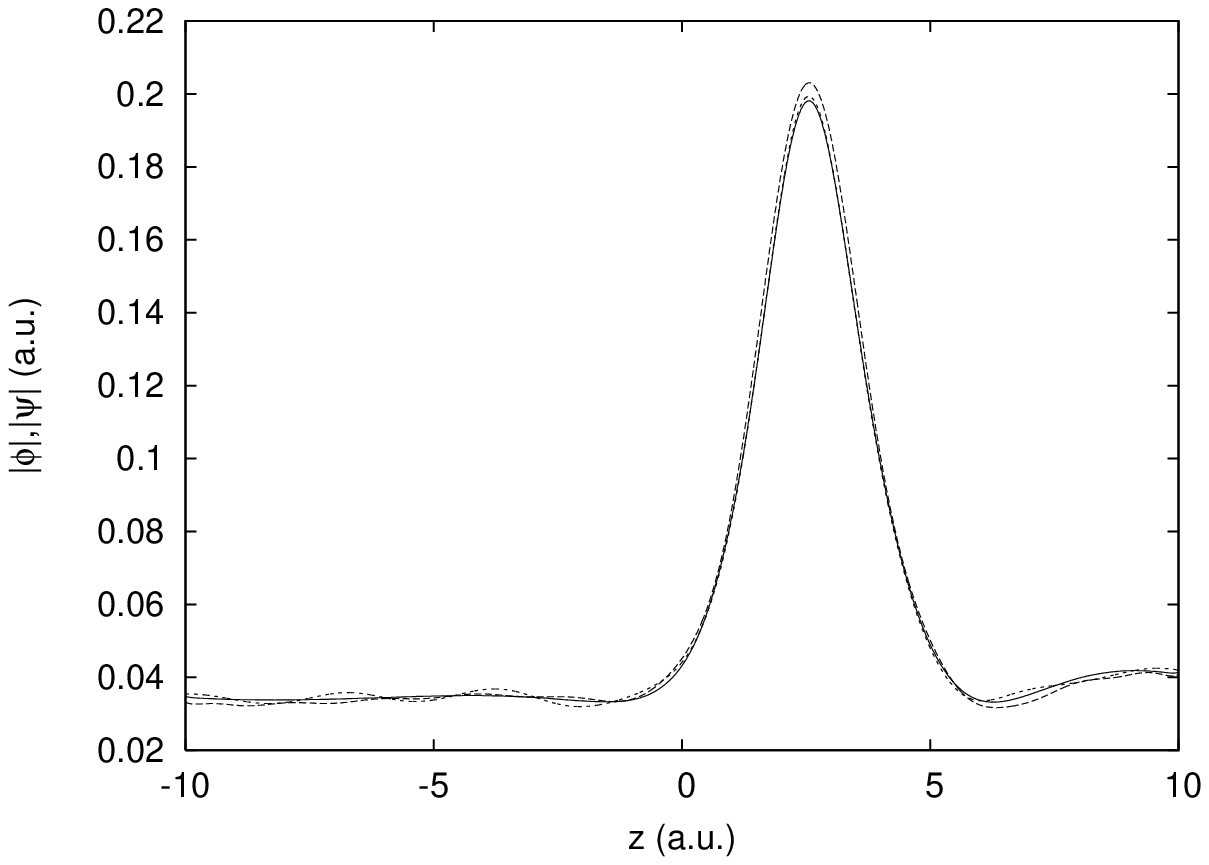}(a)
\includegraphics*[width=12cm]{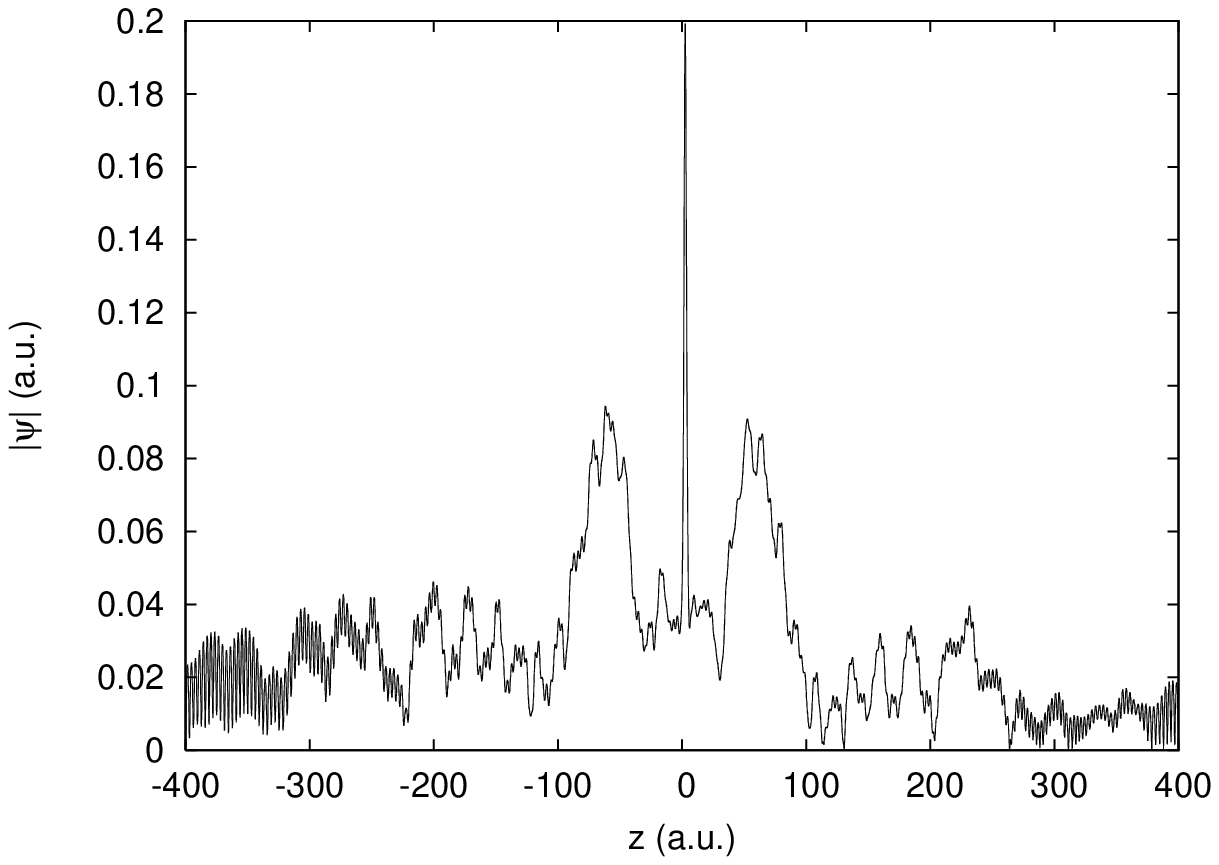}(b)
\caption{Oscillating potential: magnitude of wave-function at $t=400$ 
a.u., 12.73 periods.
\newline
(a) $|\tilde{\phi}|$ calculated using embedding: solid line, 40 basis
functions; dashed line, 25 basis functions. $|\tilde{\psi}|$
calculated over extended space using finite differences: short-dashed
line. Plotted over embedding region between $z=\pm 10$ a.u.
\newline(b) $|\tilde{\psi}|$ plotted over extended region between
$z=\pm 400$ a.u., showing reflection of the wave-function at the
boundaries of the finite-difference calculation, and subsequent
interference.}
\label{fig7} 
\end{figure}

Figures \ref{fig5}, \ref{fig6} and \ref{fig7} show the comparison
between the embedded and finite difference wave-functions in region I,
for times $t=80$ a.u., 320 a.u., and 400 a.u. (the period of the
oscillating potential is $31.42$ a.u.) and we see that the agreement
is very good. At the scale of figures \ref{fig5}(a)-\ref{fig6}(a), the
magnitude of the embedded wave-function $|\tilde{\phi}|$, calculated
with 40 basis functions, is indistinguishable from the magnitude of
the extended wave-function $|\tilde{\psi}|$. From figures
\ref{fig5}(b)-\ref{fig6}(b), we see that the error in $|\tilde{\phi}|$
is about $5\times 10^{-4}$ a.u., with a slight decrease in accuracy
with increasing time. The results with 25 basis functions are less
accurate -- they are just distinguishable from the extended results on
the scale of figures \ref{fig5}(a) and \ref{fig6}(a). At $t=400$ a.u.
$|\tilde{\psi}|$ shows slight oscillations about the embedded
$|\tilde{\phi}|$ with 40 basis functions (figure \ref{fig7}(a)). The
reason for this is that the extended wave-function has been reflected
from the limits of its range at $z=\pm 400$ a.u., with resulting
interference, as we can see from figure \ref{fig7}(b). Our embedding
results are less accurate than those of Boucke \emph{et al.} 
\cite{boucke}, who achieve a relative accuracy in their embedded
wave-function of about $5\times 10^{-5}$; however, they apply their
embedding as a boundary condition on a much larger finite-difference
calculation.

\subsection{Surface excitation in a field}
\begin{figure}[h]
\begin{center}
\epsfig{width=12cm,file=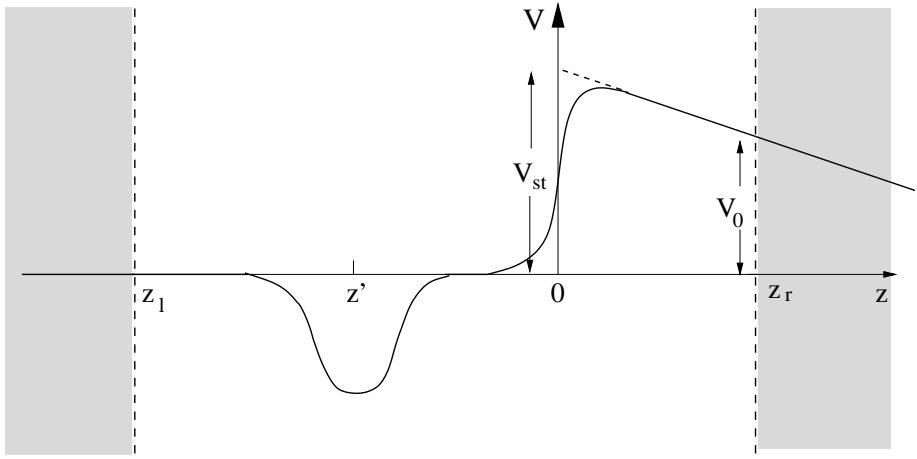}
\end{center}
\caption{Model atomic potential near a jellium surface, with applied
electric field in the vacuum. Region I, treated explicitly, lies
within embedding boundaries at $z_l$ and $z_r$.}
\label{fig8}
\end{figure}
In this section we shall follow the time-evolution of an electron
wave-function, initially in the $1/\sqrt 2\cosh(z-z')$ bound state of
a $-1/\cosh^2 (z-z')$ potential near a surface, at which there is a
constant applied electric field $\mathcal{E}$ (figure \ref{fig8}). The
surface potential step $V_{\mbox{\tiny st}}$ is broadened, so the
static potential felt by the electron is given by
\begin{equation}
V(z)=-\frac{1}{\cosh^2 (z-z')}+V_{\mbox{\tiny st}}
\left(\frac{1+\tanh z/\xi}{2}\right)-\mathcal{E}z\theta(z),
\label{surf1}
\end{equation}
where $\theta(z)$ is the step function. The surface parameters we use
are $V_{\mbox{\tiny st}}=0.5$ a.u., $\xi=0.5$ a.u., and we position
the ``atomic'' potential at $z'=-4$ a.u. The electric field is
$\mathcal{E}=0.2$ a.u., approximately $10^{11}$ Vm$^{-1}$ -- this is
much larger than the fields used in field emission but it could
represent the field of an IR laser in a quasi-static approximation.
To excite the bound state electron, an additional time-dependent
potential is applied, of the form
\begin{equation} 
\delta V(z,t)=a\exp-\left(\frac{z-z'}{\zeta}\right)^2
\sin(\omega t),
\label{surf2}
\end{equation}
with $a=1$ a.u., $\zeta=2$ a.u., and frequency $\omega=1$ a.u. This
is turned on at $t=0$, and we follow the subsequent time-evolution of
the bound-state electron wave-function. 

Region I lies in the surface region between the embedding boundaries
at $z_l$ and $z_r$ (figure \ref{fig8}). We take $z_l=-14$ a.u., and
$z_r=6$ a.u., so the embedding region extends $\pm 10$ a.u. on either
side of the atomic potential. The Hamiltonian for region I is embedded
on to the zero-potential embedding kernel (\ref{td4}) at $z_l$; at
$z_r$ it is embedded on to the Airy function kernel evaluated
numerically (section 3.2), shifted by the constant potential $V_0$
(figure \ref{fig8}), using the shift formula (\ref{td18}). We use the
basis functions given by (\ref{try2}) to expand the time-dependent
wave-function in region I -- these are centred on the atomic potential
with $D=13$ a.u.

\begin{figure}
\centering
\includegraphics*[width=12cm]{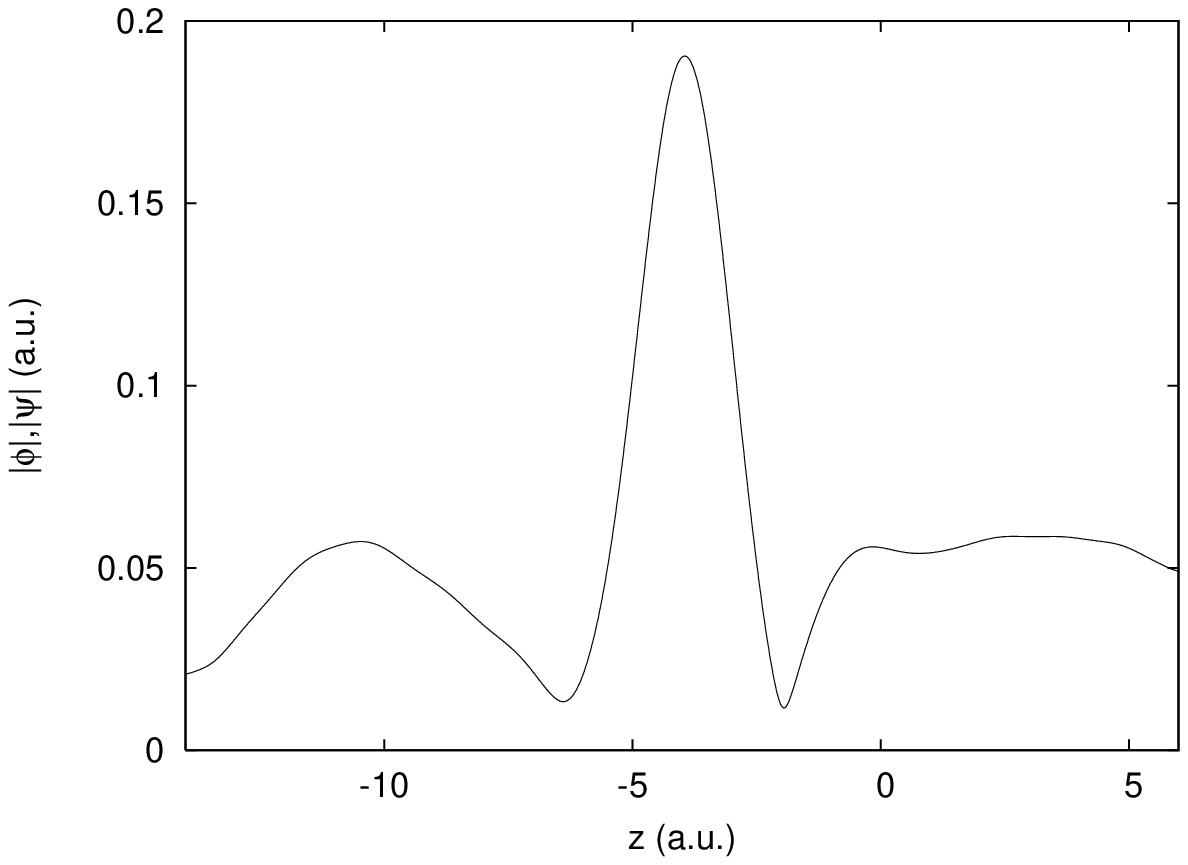}(a)
\includegraphics*[width=12cm]{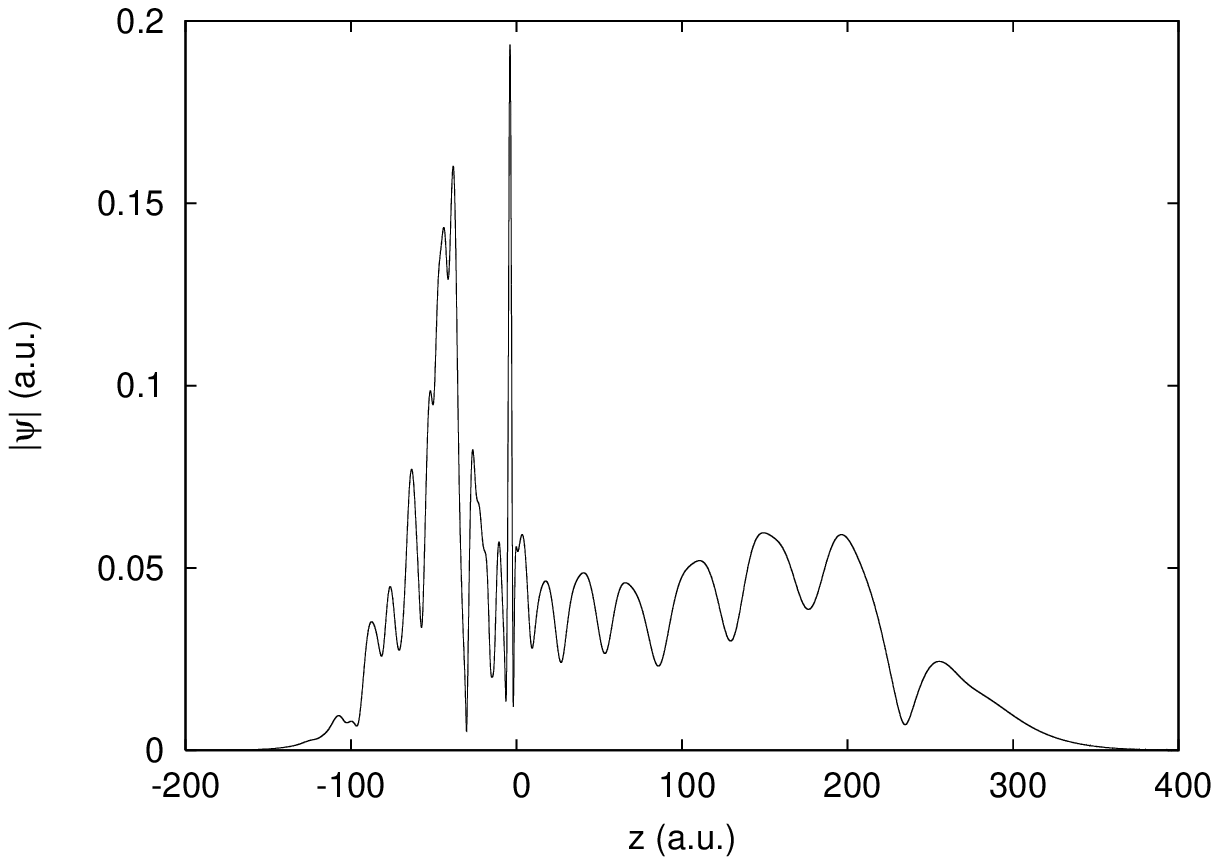}(b)
\caption{Excitation in a field, $\mathcal{E}=0.2$ a.u.: magnitude of 
wave-function at $t=50$ a.u.
\newline
(a) $|\tilde{\phi}|$ calculated using embedding with 25 basis
functions, solid line; $|\tilde{\psi}|$ calculated over extended space
using finite differences, dashed line. Plotted over embedding
region between $z=-14$ a.u. and 6 a.u.
\newline(b) $|\tilde{\psi}|$ plotted over extended region.}
\label{fig9}
\end{figure}
\begin{figure}
\centering
\includegraphics*[width=12cm]{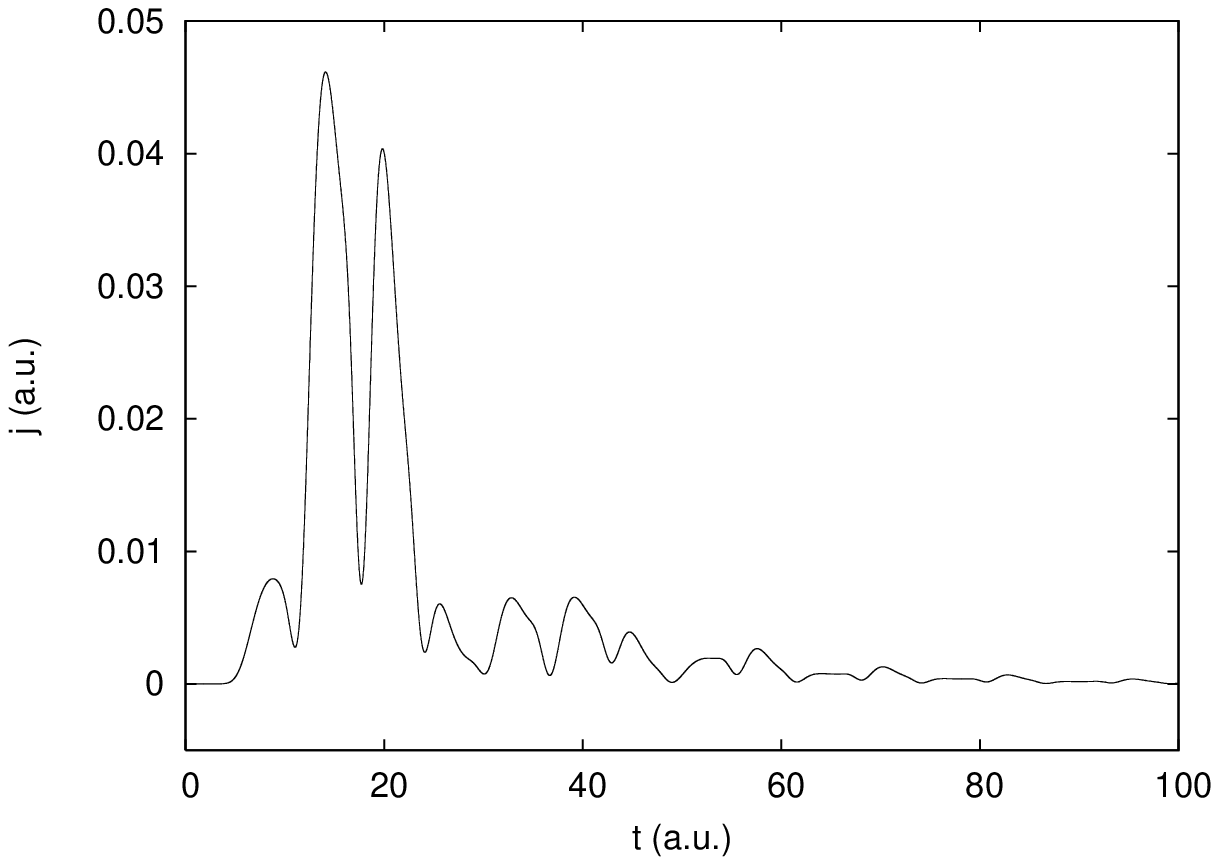}(a)
\includegraphics*[width=12cm]{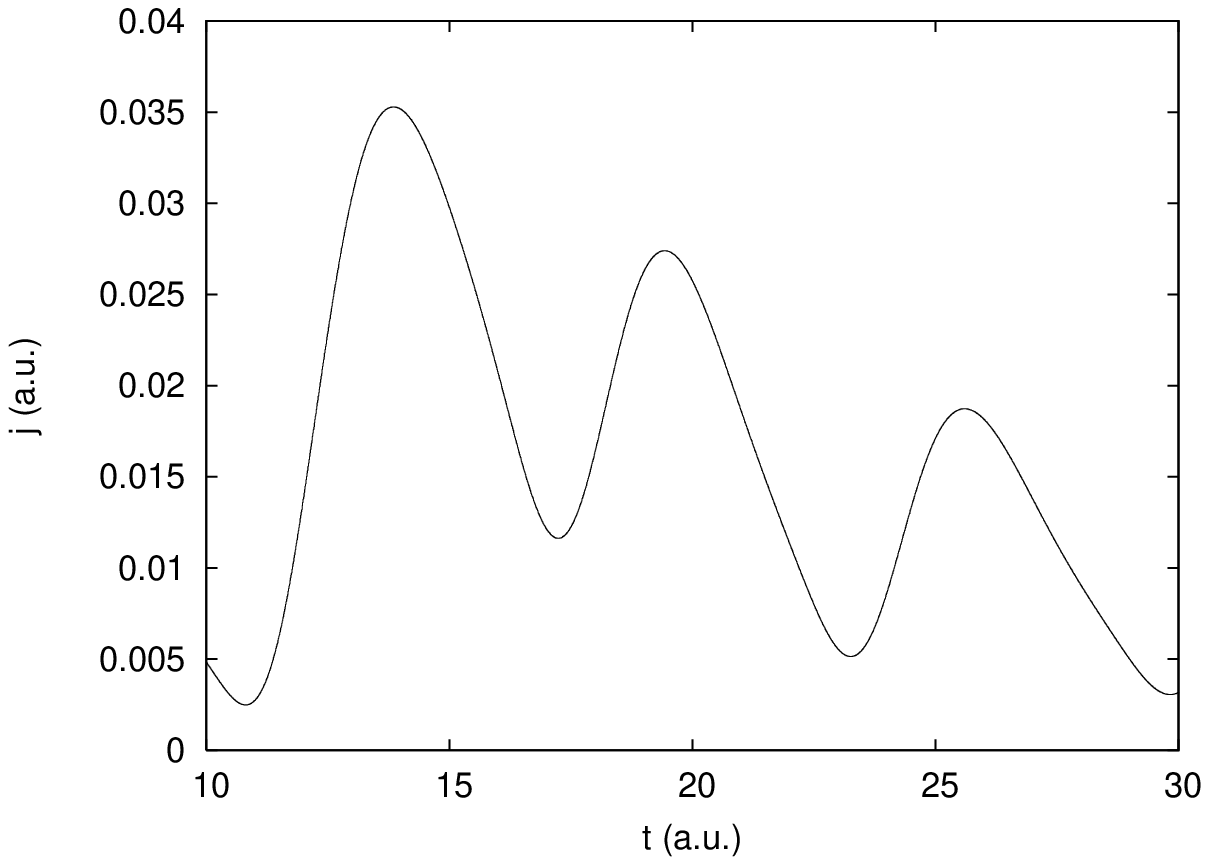}(b)
\caption{Excitation in a field, $\mathcal{E}=0.2$ a.u.: current 
across boundaries of region I as a function of time.
\newline
(a) $j$ calculated with embedding: solid line, across left-hand
boundary at $z_l$; dashed line, across right-hand boundary at $z_r$.
\newline(b) Comparison of $j$ at $z_r$ using embedding (solid line),
with finite differences (dashed line).} 
\label{fig10}
\end{figure} 

Results for $\tilde{\phi}(z,t)$ at $t=50$ a.u. are shown in figure
\ref{fig9}(a), calculated with 25 basis functions, compared with a 
finite-difference calculation for $\tilde{\psi}(z,t)$ calculated over
an extended range, $-400<z<400$ a.u. (figure \ref{fig9}(b)). We see
that agreement is very satisfactory, though not quite as good as in
section 4.1 where we used an analytic embedding kernel. In this
calculation we need to take $\delta t =0.0025$ a.u. in the time
integration, compared with 0.01 a.u. in section 4.1. This is probably
because of the difference in the time-dependent perturbation.

We also calculate the current crossing the embedding boundaries at
$z_l$ and $z_r$, using the expression
\begin{equation}
j(z,t)=\mathcal{I}\mbox{m}\left(\tilde{\phi}^*\frac{\partial
\tilde{\phi}}{\partial z}\right),
\label{surf3}
\end{equation}
with the normal derivative determined from the embedding formula
(\ref{emb18}). Taking the derivative outward from region I, a positive
current indicates charge leaving the region. Our results are shown in
figure \ref{fig10}(a) for the current crossing each boundary as a
function of time -- here we have something physical, which could in
principle be compared with experiment. There is excellent agreement
with the current calculated from the finite difference results, as we
see from figure \ref{fig10}(b), giving the current in the time range
where there is the biggest difference between the two methods.

\section{Time-evolution of extended states}
The formalism developed up to now assumes that the wave-function
$\tilde{\phi}(\mathbf{r},t)$, whose time evolution we study in region
I, has zero amplitude for $t<0$ at the embedding surface and in region
II. But in condensed matter applications we are usually interested in
exciting bulk states to which this condition does not apply, and to
study their time-evolution we have to extend the formalism.

We start with a wave-function $\Psi(\mathbf{r})$ which is an
eigenstate with energy $E$ of the time-independent Hamiltonian $H_0$,
extending through regions I and II. For times $t>0$ a time-dependent
perturbing potential $\delta V(\mathbf{r},t)$ is applied -- confined
to region I -- and the wave-function is subsequently given throughout
space by
\begin{equation}
\tilde{\psi}(\mathbf{r},t)=\Psi(\mathbf{r})\exp(-iEt)+
\tilde{\eta}(\mathbf{r},t).
\label{ext1}
\end{equation}
Substituting into the time-dependent Schr\"odinger equation gives
\begin{equation}
[H_0+\delta V(t)]\tilde{\eta}(t)+\delta V(t)\Psi\exp(-iEt)=
i\frac{\partial\tilde{\eta}}{\partial t}
\label{ext3}
\end{equation}
-- $\tilde{\eta}(\mathbf{r},t)$ satisfies the time-dependent
Schr\"odinger equation with an additional inhomogeneous term. In
region II, where $\delta V=0$, this term vanishes and $\tilde{\eta}$
satisfies the original Schr\"odinger equation,
\begin{equation}
H_0\tilde{\eta}(t)=i\frac{\partial\tilde{\eta}}{\partial t},
\label{ext4}
\end{equation}
so $\tilde{\eta}$ and its normal derivative across S are related by
the Dirichlet-to-Neumann result (\ref{emb18}),
\begin{equation}
\frac{\partial\tilde{\eta}(\mathbf{r}_S,t)}{\partial n_S}=-2
\int_S d\mathbf{r}_S\int_0^t dt'\bar{G}_0^{-1}(\mathbf{r}_S,
\mathbf{r}'_S;t-t')\frac{\partial\tilde{\eta}(\mathbf{r}_S',t')}
{\partial t'}.
\label{ext5}
\end{equation}
This means that we can write a time-dependent Schr\"odinger equation
for $\tilde{\eta}(\mathbf{r},t)$ analogous to (\ref{emb242}), with the
extra inhomogeneous term,
\begin{eqnarray}
\lefteqn{\left(-\frac{1}{2}\nabla^2+V_0(\mathbf{r})
+\delta V(\mathbf{r},t)\right)\tilde{\eta}(\mathbf{r},t) 
+\delta V(\mathbf{r},t)\Psi(\mathbf{r})\exp(-iEt)}\hspace{12
cm}\nonumber\\
+\delta(\mathbf{r}-\mathbf{r}_S)\left[\frac{1}{2}
\frac{\partial\tilde{\eta}}{\partial n_S}+
\int_S d\mathbf{r}_S'\int_0^t dt'\bar{G}_0^{-1}(\mathbf{r}_S,
\mathbf{r}'_S;t-t')\frac{\partial\tilde{\eta}(\mathbf{r}_S',t')}
{\partial t'}\right]=i\frac{\partial\tilde{\eta}}{\partial t}.
\label{ext6}
\end{eqnarray}
Solving this equation within region I gives us the change in bulk
wave-function. Note that in this section we use
$\tilde{\eta}(\mathbf{r},t)$ for the change in wave-function with
$\mathbf{r}$ in region I, evaluated by embedding (analogous to
$\tilde{\phi}$ in previous sections), as well as the change in
wave-function extended through all space (analogous to
$\tilde{\psi}$).

As a model problem we consider the jellium surface with a smeared-out
potential step,
\begin{equation}
V_0(z)=V_{\mbox{\tiny st}}\left(\frac{1+\tanh z/\xi}{2}\right),
\label{ext7}
\end{equation}
using the same parameters as in section 4.2, $V_{\mbox{\tiny st}}=0.5$
a.u., $\xi=0.5$ a.u. Region I lies within embedding boundaries at
$z_l=-10$ a.u.  and $z_r=10$ a.u., beyond which the potential is taken
as constant. The bulk continuum wave-function is found numerically
using Numerov's method \cite{jos}, matching $\Psi(z)$ on to the
asymptotic solutions of the time-independent Schr\"odinger equation
\begin{equation}
\Psi(z)=\left\{\begin{array}{l}\sin(kz+\phi),\;\;\;z<z_l\\
\alpha\exp(-\gamma z),\;\;\;z>z_r \end{array}\right.,
\label{ext8}
\end{equation}
with
\begin{equation}
k=\sqrt{2E},\;\;\gamma=\sqrt{(2(V_{\mbox{\tiny st}}-E)}.
\label{ext9}
\end{equation}
We use the same time-dependent perturbing potential as in section 4.2
(\ref{surf2}), but in this case centred at the surface step. Region I
is embedded at $z_l$ on to the free-electron embedding potential given
by (\ref{td4}), and to the right on to this embedding potential
shifted by the potential step $V_{\mbox{\tiny st}}$, following
(\ref{td18}).

The expansion coefficients in the basis set expansion of $\tilde{\eta}
(z,t)$ in region I satisfy (\ref{try5}) and (\ref{try7}), with an
extra term $b_i$ added on to $\Gamma_i$ (\ref{try8}) given by
\begin{equation}
b_i=\exp(-iEt)\int_{z_l}^{z_r}dz \chi_i(z)\delta V(z,t)\Psi(z).
\label{ext10}
\end{equation}
As in the surface electric field calculation in section 4.2, we take
$\delta t=0.0025$ a.u. in the time-integration for accurate results.

\begin{figure}
\centering
\includegraphics*[width=12cm]{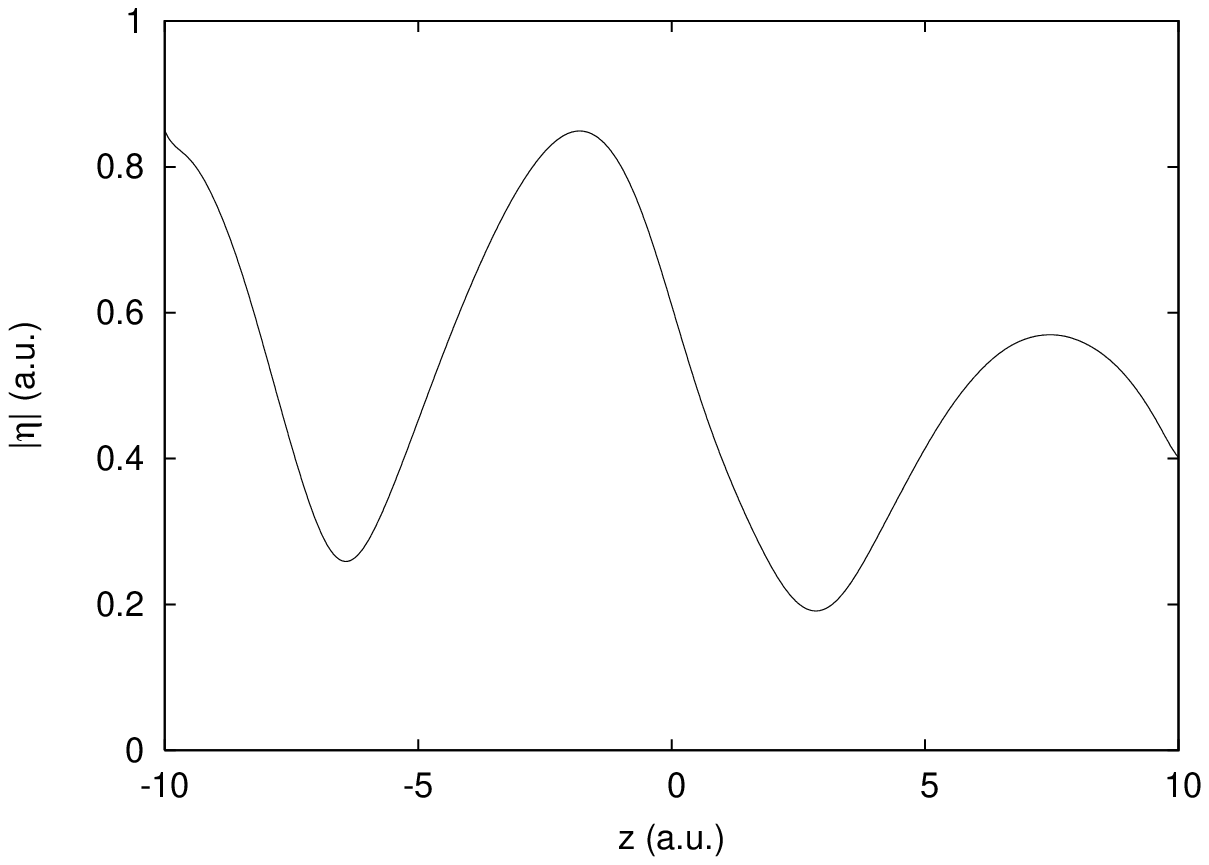}(a)
\includegraphics*[width=12cm]{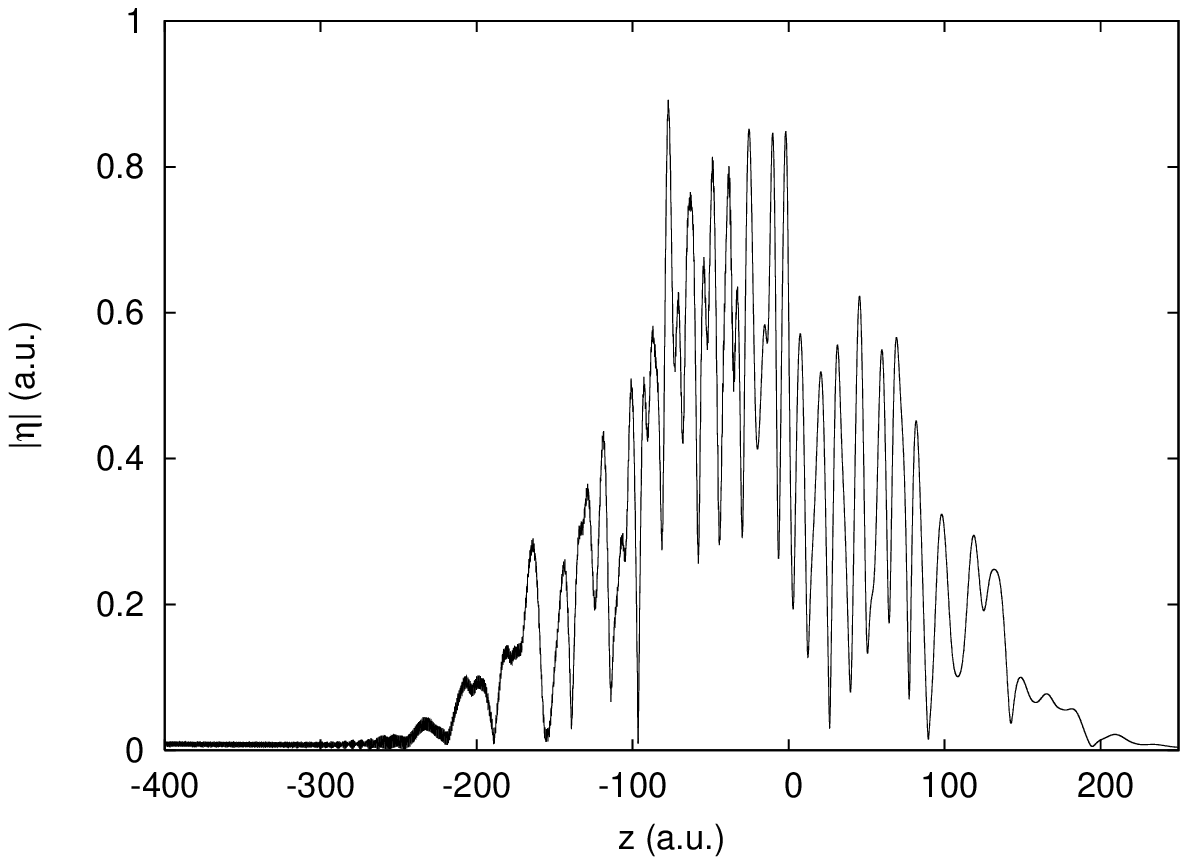}(b)
\caption{Excitation of a bulk state: magnitude of change in 
wave-function at $t=120$ a.u.
\newline
(a) $|\tilde{\eta}|$ calculated using embedding with 25 basis
functions, solid line; $|\tilde{\eta}|$ calculated over extended space
using finite differences, dashed line. Plotted over embedding
region between $z=-10$ a.u. and 10 a.u.
\newline(b) $|\tilde{\eta}|$ from finite differences plotted over 
extended region.}
\label{fig11}
\end{figure}
\begin{figure}
\centering
\includegraphics*[width=12cm]{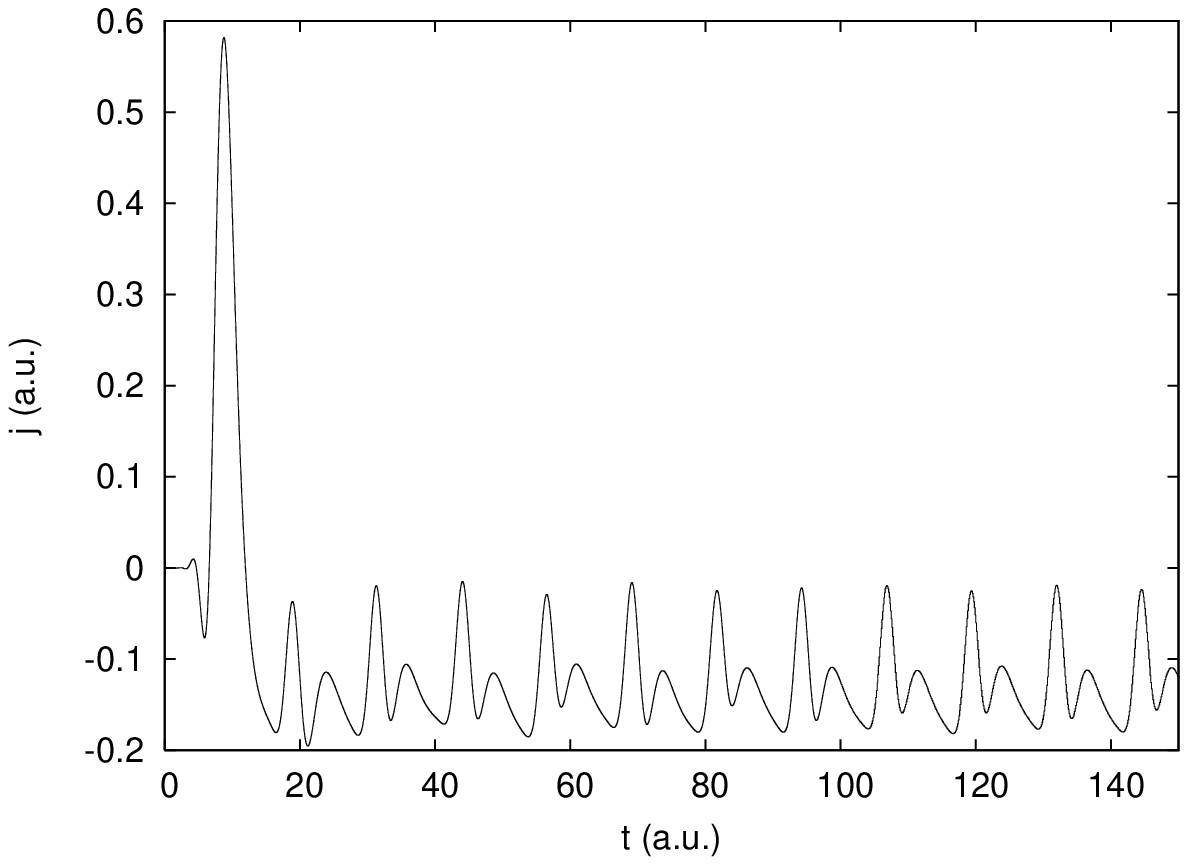}(a)
\includegraphics*[width=12cm]{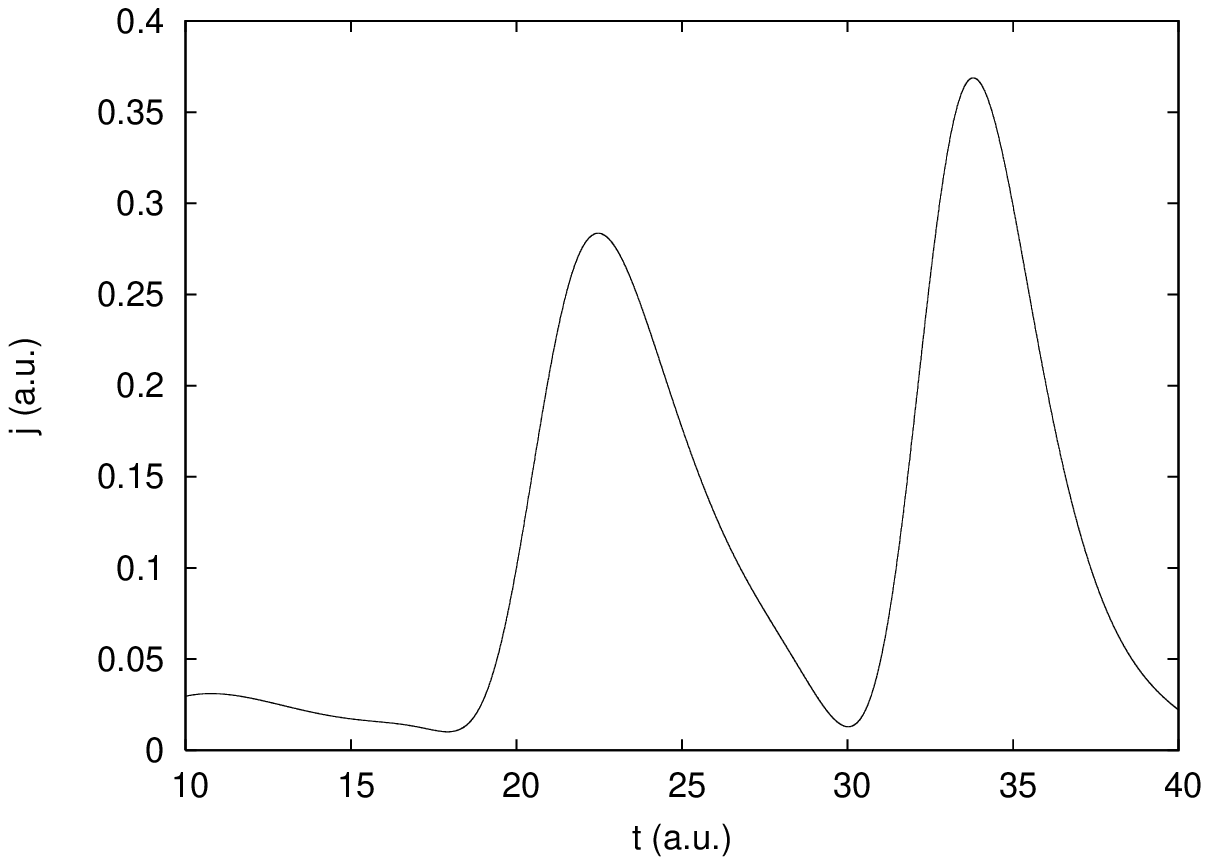}(b)
\caption{Excitation of a bulk state: current across boundaries of 
region I as a function of time. Positive current corresponds to charge
\emph{leaving} region I.
\newline
(a) $j$ calculated with embedding: solid line, across left-hand
boundary at $z_l$; dashed line, across right-hand boundary at $z_r$.
\newline(b) Comparison of $j$ at $z_r$ using embedding (solid line),
with finite differences (dashed line).}
\label{fig12} 
\end{figure}

We take a bulk state with energy $E=0.3$ a.u. and apply the
time-dependent surface perturbation with frequency $\omega=0.5$ a.u.,
starting at $t=0$. The results for the modulus of the change in
wave-function at $t=120$ a.u., calculated with 25 basis functions, are
shown in figure \ref{fig11}(a), compared with results from a
finite-difference calculation taken over the extended range
$-1000<z<1000$ a.u. Figure \ref{fig11}(b) shows the finite difference
results over part of the extended range, and we see edge effects,
propagating from the left-hand boundary and reaching $z=-100$
a.u. With the left-hand boundary in the finite difference calculation
at $-1000$ a.u., the results in the surface region become unusable
beyond $t=120$ a.u., a problem which of course does not affect
embedding with its correct treatment of boundary conditions. We see
from figure \ref{fig11}(a) that the embedding results are accurate,
even with this relatively small basis set.

\begin{figure}
\centering
\includegraphics*[width=11cm]{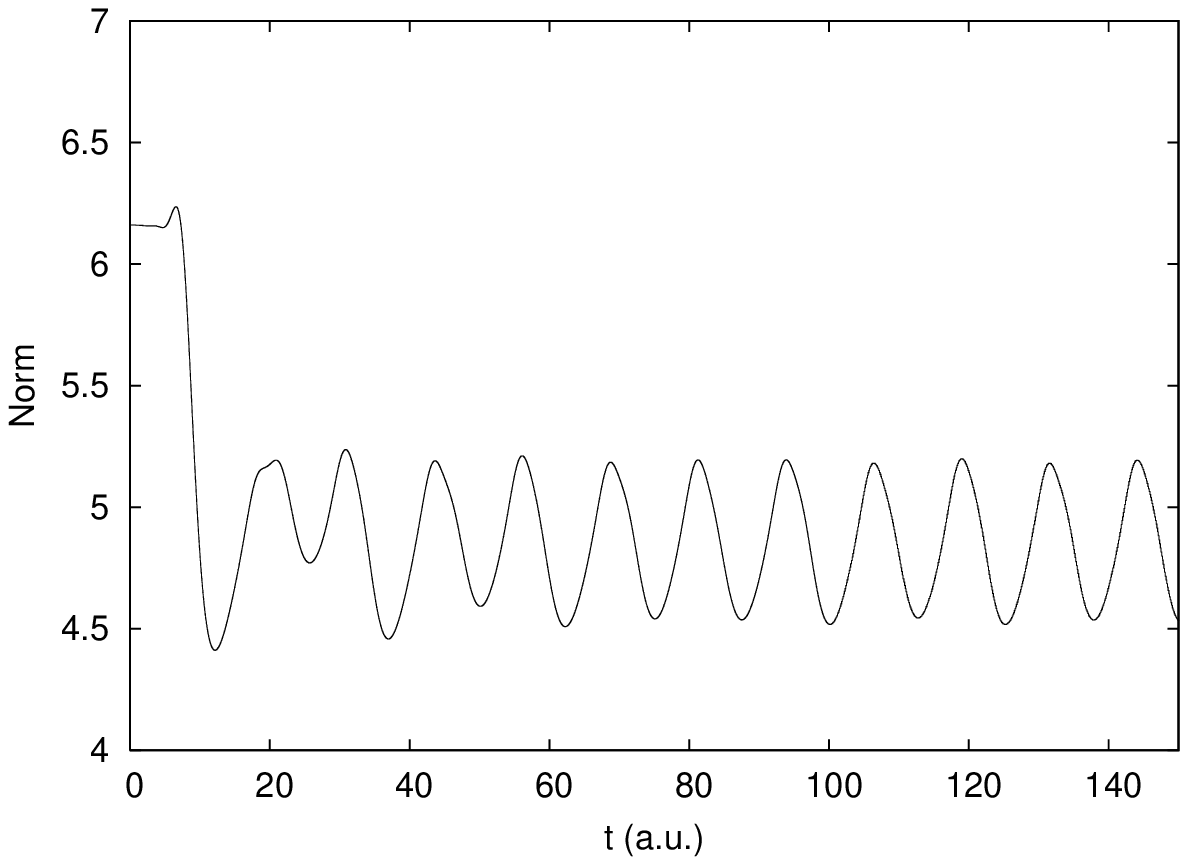}
\caption{Norm of wave-function in the embedding region as a function
of time.}
\label{fig13}
\end{figure}

It is interesting to calculate the current in this case, not only as a
sensitive test of the accuracy of the calculation, but also to
illustrate the physics. Of course we must use the full wave-function
given by (\ref{ext1}) in the expression (\ref{surf3}) for the
current. Checking the accuracy first, we see from figure
\ref{fig12}(b) that embedding works well. It is figure 
\ref{fig12}(a) which has physical content, and we see that after a 
short period of transient behaviour, the current entering the surface
from the bulk (a negative current at $z_l$) and leaving the surface
into vacuum (a positive current at $z_r$) both settle down to steady
behaviour. In fact the currents balance out on average, as is shown by
figure \ref{fig13}, giving the norm of $\tilde{\psi}(z,t)$ in the
embedding region. We see that the charge in the surface region
oscillates about a constant value, after a sudden loss of charge at
about $t=10$ a.u.

The results shown in figure \ref{fig12}(a) are in some ways surprising
-- we are solving the Schr\"odinger equation in the surface region
with an embedding potential based on an outgoing Green function, and
yet we are able to describe the current \emph{entering} the surface
from the bulk. This goes to show the power of Green function-based
methods.

\section{Conclusions}
This embedding method provides the correct boundary conditions for
solving the time-dependent Schr\"odinger equation in a limited region
of space, region I, automatically matching the solution on to the
time-evolving wave-function in the rest of the system, region II. Once
we have found the embedding kernel for region II, all that we have to
do is to add this on to the Hamiltonian for region I and
time-integrate the Schr\"odinger equation in this region.  As we solve
the time-dependent Schr\"odinger equation using the relatively small
basis set needed to describe region I, this embedding method is very
economical. In the examples in this paper we use a plane wave basis to
expand the the wave-function in region I, but any other basis set
(see, for example ref. \cite{jos2}) could be used.

The next stage in this project is to improve the numerical
time-integration scheme, and then apply it to more realistic surface
models to study surface electron excitation.

\section*{Acknowledgements}
The author would like to thank Hiroshi Ishida, Pedro Echenique, and
Txema Pitarke for their hospitality. It was during stays at their
institutions that most of this work was carried out.

\end{document}